\magnification=1200
\def\qed{\unskip\kern 6pt\penalty 500\raise -2pt\hbox
{\vrule\vbox to 10pt{\hrule width 4pt\vfill\hrule}\vrule}}
\null\bigskip\bigskip
\centerline{TOPICS IN QUANTUM STATISTICAL MECHANICS}
\centerline{AND OPERATOR ALGEBRAS.}
\bigskip
\centerline{by David Ruelle\footnote{*}{These notes correspond to lectures given at Rutgers, Math. Dept. in 1999-2000.  Email address $<$ruelle@ihes.fr$>$}.}
\bigskip\bigskip\noindent
	{\sl Abstract.  The language of operator algebras is of great help for the formulation of questions and answers in quantum statistical mechanics.  In Chapter 1 we present a minimal mathematical introduction to operator algebras, with physical applications in mind.  In Chapter 2 we study some questions related to the quantum statistical mechanics of spin systems, with particular attention to the time evolution of infinite systems.  The basic reference for these two chapters is Bratteli-Robinson: Operator algebras and quantum statistical mechanics I, II.  In Chapter 3 we discuss the nonequilibrium statistical mechanics of quantum spin systems, as it is currently being developped.}
\vfill\eject
\null\bigskip\bigskip
\centerline{Chapter 1. ALGEBRAS OF OPERATORS IN HILBERT SPACE\footnote{*}{The result presented in this chapter are classical and listed without proofs.  For the convenience of the reader some references are given to Bratteli-Robinson [1] and Reed-Simon [2] vol I.}.}
\bigskip\bigskip
	{\bf 1. Standard facts on Banach spaces, Hilbert spaces and operators.}
\medskip
	For definiteness we recall here a certain number of definitions and results, to which the reader may refer as needed.
\medskip
	{\sl Normed and Banach spaces.}
\medskip
	We shall use vector spaces over the field ${\bf R}$ or ${\bf C}$ of {\it scalars}, and extend from ${\bf R}$ to ${\bf C}$ by the replacement $E\to E\otimes_{\bf R}{\bf C}$ when useful.  A function $||.||$ on $E$ is a {\it norm} if $||\lambda x||=|\lambda| ||x||$ when $\lambda$ is a scalar, $||x+y||\le||x||+||y||$, and $||x||=0$ implies $x=0$.  In the {\it normed space} $(E,||.||)$, a set of the form $B_a(r)=\{x:||x-a||<r\}$, with $r\ge0$, is called an {\it open ball} with center $a$ and radius $r$; any union of open balls is an {\it open set}, and these open sets define the topology of the normed space.  We say that $(E,||.||)$ is {\it separable} if there is a countable set $S\subset E$ such that $S$ is dense in $E$.  A sequence $(x_n)$ in $E$ is {\it Cauchy} if $\lim_{m,n\to\infty}||x_n-x_m||=0$.  We say that $(E,||.||)$ is complete with respect to the norm, or is a {\it Banach space} if every Cauchy sequence is convergent ({\it i.e.}, $\lim_{a\to\infty}||x_n-x||=0$ for some $x\in E$, called the limit of $(x_n)$).
\medskip
	{\sl Linear functionals and operators.}
\medskip
	If $(E,||.||)$, $(F,||.||)$ are normed spaces, the linear function $A:E\to F$ is continuous if and only if it is bounded, {\it i.e.}, 
$$	||A||=\sup_{x\ne0}{||Ax||\over||x||}<0\eqno{(1)}      $$
The bounded linear functions $A:E\to F$ form a Banach space with respect to the norm\footnote{**}{The norms on $E$, $F$, and bounded linear maps $E\to F$ are in general different, although we denote them all by $||.||$.} $A\to||A||$ defined by (1).  When $F$ is the field of scalars (${\bf R}$ or ${\bf C}$) the bounded linear $A:E\to\hbox{scalars}$ are called bounded {\it linear functionals}, and they constitute the {\it dual} of $(E,||.||)$; the dual is thus a Banach space $(E',||.||)$.  When $F=E$ (with the same norm), the bounded linear $A:E\to E$ are called bounded {\it (linear) operators} on $E$, and form a Banach space ${\cal L}$.  In fact, ${\cal L}$ is a Banach algebra (see Section 3) with unit ${\bf 1}$ (the identity operator on $E$).
\medskip
	The $w^*$-topology on the dual $E'$ is the topology of pointwise convergence of functions on $E$ ({\it i.e.}, if $\xi,\xi_n\in E'$ we write $\xi_n\to\xi$ when $\xi_n(x)\to\xi(x)$ for every $x\in E$).  The unit ball $\{\xi\in E':||\xi||\le1\}$ of the dual is compact for the $w^*$-topology ({\it theorem of Alaoglu-Bourbaki}).
\medskip
	{\sl Hilbert spaces and operators in Hilbert spaces.}
\medskip
	Let again $E$ be a linear space on the field ${\bf R}$ or ${\bf C}$ of scalars.  A map $(.\,,.):E\times E\to\hbox{scalars}$ is called {\it inner product} if it is linear in the second argument, and $(x,y)=\overline{(y,x)}$ (complex conjugate), and $(x,x)\ge0$.  In particular $(.\,,.)$ is antilinear in the first argument: $(\lambda x,y)=\overline{\lambda}\,(x,y)$.  Write $||x||=(x,x)^{1/2}$; if $(E,||.||)$ is a Banach space we say that $(E,(.\,,.))$ is a {\it Hilbert space}.  A family $(a_i)_{i\in I}$ of elements of $E$ is an (orthonormal) {\it basis} if $(a_i,a_j)=\delta_{ij}$ and if the linear combinations of the $a_i$ are dense in $E$.  We call $\hbox{card}\,I$ the dimension of the Hilbert space $E$.  We shall be mostly interested in separable Hilbert spaces, which are those for which $I$ is countable (finite or infinite).  If $x\in E$, then $(x,.)$ is an element of the dual $E'$ of $E$, and $x\mapsto(x,.)$ is an antilinear isometric bijection $E\to E'$.  If the bounded operator $A$ satisfies $(x,Ax)\ge0$ for all $x\in E$, we write $A\ge0$, and say that $A$ is {\it positive}.  If $A$ is a bounded operator on the Hilbert space $(E,(.\,,.))$, there is another bounded operator $A^*$, called the {\it adjoint}, such that $(A^*x,y)=(x,Ay)$.  The map $A\mapsto A^*$ is antilinear, isometric, and an involution ({\it i.e.} $A^{**}=A$) of ${\cal L}$.  Furthermore $(AB)^*=B^*A^*$, $A^*A\ge0$, and $||A^*A||=||A||^2$.
\medskip
	The bounded operator $A$ is said to be {\it normal} if $AA^*=A^*A$, {\it self-adjoint} if $A^*=A$, and a {\it projection} if $A^*=A=A^2$.  A projection is a positive operator, a positive operator is self-adjoint, and a self-adjoint operator is normal.  The operator $U$ is said to be {\it unitary} if $UU^*=U^*U={\bf 1}$.
\medskip
	{\sl Spectral theorem.}
\medskip
	Extend if necessary the field of scalars to ${\bf C}$; for a normal operator $A$ we have the {\it spectral decomposition}
$$	A=\int\lambda{\bf P}(d\lambda)      $$
with which we assume the reader to be familiar: ${\bf P}$ is a projection-valued measure with support $S_A$ (the spectrum of $A$) such that ${\bf P}(\emptyset)=0$, ${\bf P}({\bf C})=1$, and ${\bf P}(X\cap Y)={\bf P}(X){\bf P}(Y)={\bf P}(Y){\bf P}(X)$ for subsets $X,Y$ of ${\bf C}$.  
\medskip
	For any bounded continuous function $\phi:{\bf C}\to{\bf C}$ one can define $\phi(A)=\int\phi(\lambda){\bf P}(d\lambda)$.
\medskip
	{\sl Operator topologies.}
\medskip
	There are several useful topologies on bounded operators in a Hilbert space $E$.  Convergence of $A$ to $A_0$ is given

\noindent
	in the {\it norm topology} by $||A-A_0||\to0$,

\noindent
	in the {\it strong operator topology} by $||Ax-A_0x||\to0$ for every $x\in E$,

\noindent
	in the {\it weak operator topology} by $|(x,Ay)-(x,A_0y)|\to0$ for every $x,y\in E$.
\medskip
	{\bf 2. Unbounded operators in Hilbert space.}
\medskip
	{\sl Antilinear operators.}
\medskip
	On a complex Hilbert space ${\cal H}$, it is useful to consider both linear and antilinear operators.  An {\it antilinear} operator $A:{\cal H}\to{\cal H}$ satisfies
$$	A(x+y)=Ax+Ay\qquad,\qquad A(\lambda x)=\overline{\lambda}\,Ax      $$
and boundedness is again given by (1).  The adjoint $A^*$ is again antilinear, and defined by 
$$	(A^*x,y)=\overline{(x,Ay)}\qquad(=(Ay,x))      $$
With this terminology we call {\it antiunitary} an antilinear operator $U$ such that $UU^*=U^*U={\bf 1}$.
\medskip
	[Note that we can always write (noncanonically!) ${\cal H}=E\otimes_{\bf R}{\bf C}$ where $E$ is a real Hilbert space, and that complex conjugation $J$ is then an antiunitary operator on ${\cal H}$.  Every antilinear operator is of the form $JA$ where $A$ is linear; this takes the mystery away from antilinear operators.]
\medskip
	{\sl Closed operators.}
\medskip
	We shall now remove the boundedness condition (1).  Consider an operator $A$ defined on a linear subspace $D(A)\subset{\cal H}$, and linear or antilinear $D(A)\to{\cal H}$.  We call $D(A)$ the {\it domain} of $A$; the {\it graph} of $A$ is 
$$	\Gamma=\{(x,Ax):x\in D(A)\}      $$
We say that $A$ is {\it densely defined} if $\hbox{closure }D(A)={\cal H}$; $A$ is {\it closed} if $\Gamma$ is closed in ${\cal H}\times{\cal H}$; $A$ is {\it closable} if $\hbox{closure }\Gamma$ is still a functional graph.  If $A$ is closable, the closure of $\Gamma$ is the graph of the smallest closed extension of $A$, called the {\it closure} of $A$.
\medskip
	If $A$ is linear and densely defined, let 
$$  D(A^*)=\{u\in{\cal H}:(\exists v)(\forall x\in D(A))(u,Ax)=(v,x)\}  $$
Then v is unique and there is a linear operator $A^*$ (called the {\it adjoint} of A) with domain $D(A^*)$ such that $A^*u=v$.  If $A$ is antilinear and densely defined, let 
$$	D(A^*)
=\{u\in{\cal H}:(\exists v)(\forall x\in D(A))(u,Ax)=\overline{(v,x)}\}      $$
Then v is unique and there is an antilinear operator $A^*$ (called the {\it adjoint} of A) with domain $D(A^*)$ such that $A^*u=v$.
\medskip
	{\bf Proposition.}
\medskip{\sl 
	Let $A$ be densely defined, then

\noindent
	(i) $A^*$ is closed,

\noindent
	(ii) $A$ is closable if and only if $A^*$ is densely defined, in which case {\rm$\hbox{closure }A=A^{**}$},

\noindent
	(iii) $A$ closable implies {\rm$(\hbox{closure }A)^*=A^*$}.}
\medskip
	See [2] Theorem VIII.1.\qed
\medskip
	The linear operator $A$ is {\it self-adjoint} if $A=A^*$, and {\it essentially self-adjoint} if $\hbox{closure }A$ is self-adjoint.  For a self-adjoint operator $A$ we have the {\it spectral decomposition theorem} (see [2] Theorem VIII.4, VIII.5, VIII.6).  In particular, for $t\in{\bf R}$, we can define $e^{iAt}$.  The family $t\mapsto U(t)=e^{iAt}$ is a {\it strongly continuous one-parameter group} of unitary operators, {\it i.e.}, $t\mapsto U(t)x$ is norm-continuous for each $x\in{\cal H}$, and  $U(s+t)=U(s)U(t)$ (in particular $U(0)=U(0)U(0)U(0)^*=U(0)U(0)^*={\bf 1}$).
\medskip
	{\bf Theorem} (Stone).
\medskip{\sl 
	If $t\mapsto U(t)$ is a strongly continuous one-parameter group of unitary operators on ${\cal H}$, then $U(t)=e^{iAt}$ with $A$ self-adjoint.}
\medskip
	See [2] Theorem VIII.8.\qed
\medskip
	In fact $A$ is unique,
$$	\hbox{graph }A=\{(x,y)\in{\cal H}\times{\cal H}:
	iy=\lim_{t\to0}{U(t)x-x\over t}\}      $$
and $A$ is called the infinitesimal generator of $t\mapsto U(t)$.  For further related results see [2].
\medskip
	A closed operator has a polar decomposition which we now describe in a special case
\medskip
	{\bf Proposition.} (Polar decomposition).
\medskip{\sl 
	Let $A$ be a closed linear (resp. antilinear) operator such that {\rm$\hbox{ker }A=\{x\in D(A): Ax=0\}={0}$} and {\rm$\hbox{im }A=AD(A)$} is dense in ${\cal H}$.  Then there are uniquely defined operators $|A|$ positive self-adjoint and $U$ unitary (resp. antiunitary) such that 
$$	A=U|A|      $$}
\medskip
	See [2] Theorem VIII.32.\qed
\medskip
	The study unbounded operators is delicate, and we refer to Reed-Simon [2] for further details.
\medskip
	{\bf 3. $B^*$-algebras and $C^*$-algebras{\rm\footnote{*}{For the results in this section see [1] Sections 2.1 and 2.2}}.}
\medskip
	{\sl Banach algebras.}
\medskip
	Let ${\cal A}$ be an associative algebra over ${\bf R}$ or ${\bf C}$ with a norm which makes it a Banach space; ${\cal A}$ is called a {\it Banach algebra} if 
$$	||AB||\le||A||\,||B||\eqno{(2)}      $$
If ${\cal A}$ has a unit element ${\bf 1}$ we assume that $||{\bf 1}||=1$.  For example, the bounded operators on a Banach space $E$ form a Banach algebra which has for unit element the identity operator on $E$.
\medskip
	{\sl Spectrum.}
\medskip
	If ${\cal A}$ is a complex Banach algebra with unit element ${\bf 1}$, and $A\in{\cal A}$ we write
$$	Sp_A=\{\lambda\in{\bf C}:
	A-\lambda{\bf 1}\hbox{ is not invertible in }{\cal A}\}      $$
and call $Sp_A$ the {\it spectrum} of $A$.  The {\it spectral radius}
$$	r_A=\sup\{|\lambda|:\lambda\in Sp_A\}      $$
satisfies the {\it spectral radius formula}
$$	r_A=\lim_{n\to\infty}||A^n||^{1/n}=\inf_n||A^n||^{1/n}      $$
\medskip
	{\sl $B^*$-algebras.}
\medskip
	An {\it adjoint operation} in an algebra over ${\bf C}$ is an antilinear map $A\mapsto A^*$ such that $A^{**}=A$ and $(AB)^*=B^*A^*$.
\medskip
	A $B^*$-algebra is a Banach algebra over ${\bf C}$ with an adjoint operation such that
$$	||A^*A||=||A||^2\eqno{(3)}      $$
[In particular (2) and (3) imply $||A^*||=||A||$].
\medskip
	If the $B^*$-algebra ${\cal A}$ has a unit element ${\bf 1}$, then ${\bf 1}^*={\bf 1}$, and $||{\bf 1}||=1$ follows from (2), (3) if we assume that $||.||$ does not vanish identically.
\medskip
	A norm-closed self-adjoint subalgebra of a $B^*$-algebra is called a {\it $B^*$-subalgebra}.  If ${\cal A}$, ${\cal B}$ are $B^*$-algebras and $\pi:{\cal A}\to{\cal B}$ a homomorphism of complex algebras such that $\pi(A^*)=(\pi A)^*$, then $\pi$ is called a {\it $*$-morphism}, or {\it morphism} of $B^*$-algebras.  The $B^*$-algebras with these morphisms form a {\it category}.
\medskip
	{\bf Proposition} (adjunction of a unit element).
\medskip
	{\sl If ${\cal A}$ is a $B^*$-algebra without unit element, one can extend the algebra structure, the norm and the adjoint operation to ${\bf C}{\bf 1}\oplus{\cal A}$ to obtain a $B^*$-algebra with unit element ${\bf 1}$.  In particular
$$	||\alpha{\bf 1}+A||
	=\sup\{||\alpha B+AB||:B\in{\cal A},||B||=1\}      $$}
\indent
	{\bf Proposition.}
\medskip
	{\sl To define the spectrum $S_A$ and spectral radius $r_A$ of $A\in{\cal A}$, we adjoin if necessary a unit element to the $B^*$-algebra ${\cal A}$.  If $A$ is normal, {\it i.e.}, $A^*A=AA^*$, then
$$	r_A=||A||      $$
For all $A\in{\cal A}$ we have thus
$$	||A||^2=||A^*A||=r_{A^*A}      $$}
\indent
	{\bf Proposition.}
\medskip
	{\sl If ${\cal B}$ is a $B^*$-subalgebra of ${\cal A}$, and $A\in{\cal B}$, then the spectrum of $A$ as an element of ${\cal B}$ is the same as its spectrum as an element of ${\cal A}$.}
\medskip
	{\bf Remark.}
\medskip
	The definition of $B^*$-algebras seems to involve a nonalgebraic element -- the norm -- but the above two propositions show that the norm is determined by the algebraic structure.  Note in this respect that morphisms have been defined without reference to the norm.
\medskip
	{\sl $C^*$-algebras.}
\medskip
	An algebra ${\cal A}$ of bounded operators on a complex Hilbert space ${\cal H}$ is called a {\it $C^*$-algebra} if it is self-adjoint (${\cal A}^*={\cal A}$) and closed for the operator norm topology.
\medskip
	{\bf Theorem.}
\medskip
	{\sl Each $C^*$-algebra is a $B^*$-algebra, and conversely each $B^*$-algebra is isomorphic to a $C^*$-algebra.}
\medskip
	In view of this result it is usual to speak of $C^*$-algebras instead of $B^*$-algebras.
\medskip
	{\sl Abelian $B^*$-algebras.}
\medskip
	Let $X$ be a locally compact space and ${\cal C}_0(X)$ the algebra of continuous functions $X\to{\bf C}$ and tending to 0 at infinity.  With respect to complex conjugation as adjoint operation and the sup-norm, ${\cal C}_0(X)$ is a commutative $B^*$-algebra.  Conversely, every commutative $B^*$-algebra ${\cal A}$ is isomorphic to an algebra ${\cal C}_0(X)$ obtained as follows:
\medskip
	{\bf Theorem} (Gel'fand isomorphism).
\medskip
	{\sl Let ${\cal A}$ be a commutative $B^*$-algebra, and $X$ the set of {\it characters}: $x:{\cal A}\to{\bf C}$ ($x$ is linear and $x(AB)=x(A)x(B)$).  Write $(\pi(A))x=x(A)$, and place on $X$ the topology generated by the open sets $\{x:(\pi(A))x\ne0\}$.  Then $X$ is locally compact (compact if and only if ${\cal A}$ has a unit element) and $\pi$ is an isomorphism ${\cal A}\to{\cal C}_0(X)$.}
\medskip
	{\bf 4. States on $B^*$-algebras and representations.}
\medskip
	{\sl Positive elements.}
\medskip
	An element $A$ of the $B^*$-algebra ${\cal A}$ is said to be {\it positive} ($A\ge0$) if one of the following conditions is satisfied:

\noindent (a) $A=A^*$ and $Sp_A\ge0$,

\noindent (b) $A=B^*B$ for some $B\in{\cal A}$.
\medskip
	[The equivalence is proved in [1] Theorem 2.2.12].
\medskip
	{\sl States.}
\medskip
	Consider a continuous linear functional $\rho$ on the $B^*$-algebra ${\cal A}$, and assume that $\rho\ge0$ ($\rho$ is {\it positive}, {\it i.e.}, $A\ge0\Rightarrow\rho(A)\ge0$).  When furthermore $||\rho||=1$ we say that $\rho$ is a {\it state}.  If ${\cal A}$ has a unit element ${\bf 1}$  and $\rho\ge0$, then $||\rho||=1$ if and only if $\rho({\bf 1})=1$.  A state on a $B^*$-algebra has a unique extension to a state on the algebra obtained by adjunction of a unit element.
\medskip
	The set $E$ of states on ${\cal A}$ is a convex subset of the dual of ${\cal A}$.  If ${\cal A}$ has a unit element, $E$ is compact for the $w^*$-topology of the dual.
\medskip
	{\sl Representations.}
\medskip
	A {\it representation} of a $B^*$-algebra ${\cal A}$ is a pair $({\cal H},\pi)$ where ${\cal H}$ is a complex Hilbert space and $\pi$ is a morphism of ${\cal A}$ to the $C^*$-algebra ${\cal L}({\cal H})$ of bounded operators on ${\cal H}$.  The representation is faithful if $\pi(A)=0$ implies $A=0$
\medskip
	A {\it cyclic representation} of ${\cal A}$ is a triple $({\cal H},\pi,\Omega)$ where $({\cal H},\pi)$ is a representation and $\Omega\in{\cal H}$ is such that $||\Omega||=1$ and $\pi({\cal A})\Omega$ is dense in ${\cal H}$.
\medskip
	{\sl Groups of automorphisms.}
\medskip
	If the morphism $g:{\cal A}\to{\cal A}$ has an inverse $g^{-1}$, it is called an ${\it automorphism}$ (or $*$-automorphism).  If $G$ is a group of automorphisms of ${\cal A}$, and if $\rho(gA)=\rho(A)$ for all $g\in G$, $A\in{\cal A}$, we say that the state $\rho$ is {\it invariant} (for the action of $G$ on ${\cal H})$.
\medskip
	{\sl The GNS construction (Gel'fand-Na\v\i mark-Segal)}.
\medskip
	If $({\cal H},\pi,\Omega)$ is a cyclic representation of the $B^*$-algebra ${\cal A}$, then 
$$	A\mapsto\rho(A)=(\Omega,\pi(A)\Omega)      $$ 
defines a state $\rho$ on ${\cal A}$.  Conversely, the GNS construction associates to any state $\rho$ on ${\cal A}$ a cyclic representation such that $\rho(.)=(\Omega,\pi(.)\Omega)$.  In view of applications, it is useful to consider the situation where a group $G$ of automorphisms acts on ${\cal A}$ (this group may be trivial).
\medskip
	{\bf Theorem.}
\medskip
	{\sl Let $G$ be a group of automorphisms of the $B^*$-algebra ${\cal A}$ and $\rho$ an invariant state.  There are then a cyclic representation $({\cal H}_\rho,\pi_\rho,\Omega_\rho)$ of ${\cal A}$ such that 
$$	\rho(.)=(\Omega_\rho,\pi_\rho(.)\Omega_\rho)      $$
and a unitary representation $U_\rho$ of $G$ in ${\cal H}$ such that 
$$	U_\rho(g)\Omega_\rho=\Omega_\rho\qquad,\qquad
	\pi_\rho(gA)=U_\rho(g)\pi_\rho(A)U_\rho(g)^{-1}      $$
for all $g\in G$, $A\in{\cal A}$.  The data ${\cal H}_\rho$, $\pi_\rho$, $\Omega_\rho$, $U_\rho$ as above are unique up to unitary equivalence.}
\medskip
	Dropping the index $\rho$ we sketch the construction of ${\cal H}$, $\pi$, $\Omega$, $U$.  We assume that ${\cal A}$ has a unit element (adjoin ${\bf 1}$ if necessary).  If we write
$$	{\cal N}=\{A\in{\cal A}:\rho(A^*A)=0\}      $$
and let $[\cdot]:{\cal A}\to{\cal A}/{\cal N}$ be the quotient map, there is a naturally defined scalar product $(\cdot,\cdot)$ on ${\cal A}/{\cal N}$ such that 
$$	([A],[B])=\rho(A^*A)      $$
The Hilbert space ${\cal H}$ is defined as completion of ${\cal A}/{\cal N}$ with respect to this scalar product.  One writes then
$$	\pi(A)[B]=[AB]      $$
$$	[{\bf 1}]=\Omega      $$
$$	U(g)[B]=[gB]      $$
and checks that the theorem holds with these definitions.\qed
\medskip
	{\sl Pure and ergodic states.}
\medskip
	The set $E_G$ of invariant states for the action of the group $G$ of automorphisms of ${\cal A}$ is convex.  Its extremal points are called {\it $G$-ergodic states}.  If $G$ is reduced to the the identity automorphism of ${\cal A}$, $E_G$ reduces to the set $E$ of all states, and its extremal points are {\it pure states}.
\medskip
	A set ${\cal R}$ of bounded operators on ${\cal H}$ is said to be {\it irreducible} if the only bounded operators commuting with ${\cal R}$ are multiples of ${\bf1}$:
$$	(AR=RA\hbox{ for all }R\in{\cal R})
	\Rightarrow(A=\lambda{\bf1}\hbox{ for some }\lambda\in{\bf C})      $$
\indent
	{\bf Proposition.}
\medskip
	{\sl The state $\rho$ is ergodic if and only if the set $\pi_\rho({\cal A})\cup U_\rho(G)$ of operators on ${\cal H}_\rho$ is irreducible.  In particular, $\rho$ is pure if and only if $\pi_\rho({\cal A})$ is irreducible.}
\medskip
	See [1] Theorem 4.3.17.
\medskip
	{\bf 5. Von Neumann algebras{\rm\footnote{*}{For the results in this section see [1] Section 2.4.}}.}
\medskip
	The very brief introduction to von Neumann algebras given here constitute a minimal preparation to the Tomita-Takesaki theory presented in the next section.
\medskip
	{\sl Commutant.}
\medskip
	Let ${\cal L}({\cal H})$ be the algebra of all bounded operators on the complex Hilbert space ${\cal H}$ and ${\bf 1}$ the identity operator on ${\cal H}$.  We write $[A,B]=AB-BA$ for $A,B\in{\cal L}({\cal H})$.  The {\it commutant} of a set ${\cal R}\subset{\cal L}({\cal H})$ is
$$	{\cal R}'
	=\{A\in{\cal L}({\cal H}):B\in{\cal R}\Rightarrow[A,B]=0\}      $$
If ${\cal R}'$ consists of the multiples of ${\bf 1}$, then ${\cal R}$ is {\it irreducible} (see above).  The set ${\cal R}''$=(${\cal R}')'$ is called the {\it bicommutant} of ${\cal R}$
\medskip
	{\sl Von Neumann algebras.}
\medskip
	A self-adjoint subalgebra ${\cal M}$ of ${\cal L}({\cal H})$ is called a von Neumann algebra if it satisfies one of the following equivalent conditions:

	(a) ${\cal M}\ni{\bf 1}$ and ${\cal M}$ is closed for the weak operator topology.

	(b) ${\cal M}\ni{\bf 1}$ and ${\cal M}$ is closed for the strong operator topology.

	(c) ${\cal M}$ is equal to its bicommutant: ${\cal M}={\cal M}''$.

\noindent
This equivalence is the {\it bicommutant theorem}.
\medskip
	In particular a von Neumann algebra is a $C^*$-algebra with unit element.  If ${\cal R}$ is a self-adjoint subset of ${\cal L}({\cal H})$, ${\cal R}'$ is a von Neumann algebra, and the bicommutant ${\cal R}''$ is the smallest von Neumann algebra containing ${\cal R}$.
\medskip
	The von Neumann algebra ${\cal M}$ is called a {\it factor} if ${\cal M}\cap{\cal M}'=\hbox{multiples of }{\bf 1}$, {\it i.e.}, if ${\cal M}\cup{\cal M}'$ is irreducible.
\medskip
	{\sl Predual.}
\medskip
	Let ${\cal M}$ be a von Neumann algebra on ${\cal H}$.  The linear functionals $\omega$ on ${\cal M}$ of the form
$$	A\mapsto\omega(A)=\sum_n(\xi_n,A\eta_n)      $$
where $\sum_n||\xi_n||^2<\infty$, $\sum_n||\eta_n||^2<\infty$, form a closed subspace ${\cal M}_*$ of the Banach dual ${\cal M}^*$ of ${\cal M}$, and ${\cal M}_*$ is called the {\it predual} of ${\cal M}$.  The dual of ${\cal M}_*$ is ${\cal M}$ in the duality
$$	(A,\omega)\in{\cal M}\times{\cal M}_*\mapsto\omega(A)      $$
\indent
In particular the predual of ${\cal L}({\cal H})$ can be canonically identified with the Banach space ${\cal T}({\cal H})$ of trace-class\footnote{*}{We use on ${\cal T}({\cal H})$ the trace norm $T\mapsto||T||_1=Tr(|T|)$ where $|T|=(T^*T)^{1/2}$ and the square root can be defined via the spectral theorem.} operators on ${\cal H}$ using the duality $(A,T)\in{\cal L}({\cal H})\times{\cal T}({\cal H})\mapsto Tr(TA)$.
\medskip
	{\sl Normal states.}
\medskip
	Since a von Neumann algebra ${\cal M}$ on ${\cal H}$ is a $C^*$-algebra (with unit element) we can define states on ${\cal M}$.  We call {\it normal states} those which belong to the predual ${\cal M}_*$.
\medskip
	{\bf Proposition.}
\medskip
	{\sl A state $\omega$ on ${\cal M}$ is normal if and only if there is a density matrix $\rho$, {\it i.e.}, a positive trace class operator on ${\cal H}$ with $Tr\rho=1$, such that}
$$	\omega(A)=Tr(\rho A)      $$
[Note that $\rho$ need not be unique].
\medskip
	{\sl Cyclic and separating vectors.}
\medskip
	Let ${\cal M}$ be a von Neumann algebra on ${\cal H}$.  Remember that the vector $\Omega\in{\cal H}$ is {\it cyclic} for ${\cal M}$ if ${\cal M}\Omega$ is dense in ${\cal H}$.
\medskip
	We also say that $\Omega$ is {\it separating} for ${\cal M}$ if $A\in{\cal M}$ and $A\Omega=0$ imply $A=0$.  One can check that
$$	\Omega\hbox{ cyclic for }{\cal M}\Leftrightarrow
	\Omega\hbox{ separating for }{\cal M}'      $$
\indent
	{\sl Faithful states.}
\medskip
	A state $\omega$ on a von Neumann algebra ${\cal M}$ is faithful if $\omega(A)>0$ whenever $0\ne A>0$.
\medskip
	{\bf Proposition.}
\medskip
	{\sl If $({\cal H}_\omega,\pi_\omega,\Omega_\omega)$ is the GNS representation associated with a normal state $\omega$ on a von Neumann algebra ${\cal M}$, then $\pi_\omega({\cal M})$ is again a von Neumann algebra.  If furthermore $\omega$ is faithful, then $\Omega_\omega$ is separating for $\pi_\omega({\cal M})$.  In particular, $\pi_\omega$ is an isomorphism.}
\medskip
	{\sl Abelian von Neumann algebras.}
\medskip
	For a commutative von Neumann algebra ${\cal M}$, we have the Gel'fand isomorphism ${\cal M}={\cal C}(X)$ with $X$ compact.  A normal state $\omega$ on ${\cal M}$ corresponds to a probability measure $\mu(d\omega)$ on $X$ and the GNS representation $({\cal H},\pi,\Omega)$ is given by ${\cal H}=L^2(X,\mu)$, $\Omega=\hbox{function 1 on }X$, $\pi(A)=\hbox{multiplication by }A$.  In particular $\pi({\cal M})=L^\infty(X,\mu)$ and if support $\mu=X$, one can identify the predual ${\cal M}_*$ with $L^1(X,\mu)$.
\medskip
	{\bf Remarks.}
\medskip
	The von Neumann algebras for which there exists a faithful normal state $\omega$ are said to be {\it $\sigma$-finite}; they are those von Neumann algebras for which every family of mutually orthogonal projections is countable (finite or infinite).  The von Neumann algebras relevant for physics are all $\sigma$-finite.  In this situation we may thus replace ${\cal M}$ by a $*$-isomorphic algebra (acting on a new Hilbert space) with a cyclic and separating vector: this will be the setup for the Tomita-Takesaki theory.
\medskip
	One may worry that $*$-isomorphic von Neumann algebras (isomorphic thus as C$^*$-algebras) are not really the same since they live in different Hilbert spaces.  Actually, $*$-isomorphic von Neumann algebras have the same predual, the same normal states, and are thus ``the same'' in a strong sense.  In particular, the $w^*$-topology of ${\cal M}$ as dual of its predual is uniquely defined: this is the so-called {\it $\sigma$-weak topology} (this topology is stronger than the weak operator topology, but equivalent on bounded subsets of ${\cal M}$ to the weak and the strong operator topologies).
\medskip
	Let us call W$^*$-{\it algebra} any B$^*$-algebra which is the Banach dual of some Banach space.  Then W$^*$-algebras turn out to correspond exactly to $*$-isomorphism classes of von Neumann algebras (see Sakai [3]).
\medskip
	{\bf 6. Tomita-Takesaki theory{\rm\footnote{*}{For the results in this section see [1] Section 2.5.}}.}
\medskip
	Let ${\cal M}$ be a von Neumann algebra on ${\cal H}$.  Assume that $\Omega$ is a cyclic and separating vector for ${\cal M}$, and therefore that $\Omega$ is also cyclic and separating for ${\cal M}'$.  If we write 
$$	S_0A\Omega=A^*\Omega\qquad{\rm for}\qquad A\in{\cal M}      $$
$$	F_0B\Omega=B^*\Omega\qquad{\rm for}\qquad B\in{\cal M}'      $$
we see that $S_0$, $F_0$ are densely defined on ${\cal H}$, antilinear, and such that $S_0^{-1}=S_0$, $F_0^{-1}=F_0$.
\medskip
	{\bf Proposition.}
\medskip
	{\sl $S_0$ and $F_0$ are closable operators, their closures $S$ and $F$ satisfy}
$$	S_0^*=F\qquad,\qquad F_0^*=S      $$
\indent
	{\bf Proposition} (the modular operator $\Delta$ and the modular conjugation $J$).
\medskip
	{\sl The polar decomposition 
$$	S=J\Delta^{1/2}      $$
defines a unique positive self-adjoint operator $\Delta$ (called {\it modular operator}) and a unique antiunitary operator $J$ (called {\it modular conjugation}).  These operators satisfy}
$$	\Delta=FS\qquad,\qquad\Delta^{-1}=SF      $$
$$	S=J\Delta^{1/2}\qquad,\qquad F=J\Delta^{-1/2}      $$
$$	J=J^*\qquad,\qquad J^2={\bf 1}      $$
$$	\Delta^{-1/2}=J\Delta^{1/2}J      $$
\indent
	{\bf Theorem} (Tomita-Takesaki).
\medskip
	{\sl With the above assumptions and notation}
$$	J{\cal M}J={\cal M}'      $$
$$	\Delta^{it}{\cal M}\Delta^{-it}
	={\cal M}\hbox{ for all }t\in{\bf R}      $$
\indent
	{\sl Modular automorphism group and modular condition.}
\medskip
	Let us return to the situation where $\omega$ is a faithful normal state on the von Neumann algebra ${\cal M}$, and let $({\cal H}_\omega,\pi_\omega,\Omega_\omega)$ be the corresponding cyclic representation and $\Delta$ the modular operator associated with $(\pi_\omega({\cal M}),\Omega_\omega)$.  In view of the above theorem there is a $\sigma$-weakly continuous one-parameter group $t\mapsto\sigma_t^\omega$ of $*$-automorphisms of ${\cal M}$ defined by
$$	\sigma_t^\omega(A)
	=\pi_\omega^{-1}(\Delta^{it}\pi_\omega(A)\Delta^{-it})      $$
This is called the {\it modular automorphism group} associated with the pair $({\cal M},\omega)$.  We note also the {\it modular condition}
$$	(\Delta^{1/2}\pi_\omega(A)\Omega_\omega,
	\Delta^{1/2}\pi_\omega(B)\Omega_\omega)
	=(J\pi_\omega(A^*)\Omega_\omega,J\pi_\omega(B^*)\Omega_\omega)      $$
$$	=(\pi_\omega(B^*)\Omega_\omega,\pi_\omega(A^*)\Omega_\omega)      $$
\indent
	The definition of the modular automorphism group can be extended to the situation where faithful normal states are replaced by the more general ``faithful semifinite normal weights''.  This extension is not needed however for applications to equilibrium statistical mechanics, where the modular automorphism group will represent time evolution, and the relation between the equilibrium state $\omega$ and the modular automorphisms will correspond to the {\it Kubo-Martin-Schwinger (KMS) condition}.
\medskip
	{\bf 7. KMS states.}
\medskip
	Let ${\cal A}$ be a B$^*$-algebra and $(\alpha^t)$ be a strongly continuous one-parameter group of automorphisms of  ${\cal A}$.  [The $\alpha^t$ are $*$-automorphisms of ${\cal A}$, with $\alpha^0=$ identity, $\alpha^s\alpha^t=\alpha^{s+t}$.  Strong continuity means that if $A\in{\cal A}$ then $t\to\alpha^tA$ is continuous ${\bf R}\to{\cal A}$ (with the norm topology of ${\cal A}$)].  An element $A$ of ${\cal A}$ is said to be {\it entire analytic} with respect to $(\alpha^t)$ if $t\mapsto\alpha^tA$ extends to an analytic function ${\bf C}\to{\cal A}$.
\medskip
	{\bf Lemma.}
\medskip
	{\sl The set ${\cal A}_\alpha$ of entire analytic elements with respect to $(\alpha^t)$ is a norm-dense $(\alpha^t)$-invariant $*$-subalgebra of ${\cal A}$.}
\medskip
	In fact, if $A\in{\cal A}$, the elements 
$$	A_n=\sqrt{n\over\pi}\int\alpha^t(A)e^{-nt^2}dt      $$
belong to ${\cal A}_\alpha$ and tend to $A$ when $n\to\infty$ (see [1] Proposition 2.5.22).\qed
\medskip
	{\bf Theorem} (KMS condition).
\medskip
	{\sl If $\rho$ is a state on ${\cal A}$ and $\beta>0$, the following conditions are equivalent:
\medskip
	(a) $\qquad\qquad\rho(A\alpha^{i\beta}(B))=\rho(BA)$

\noindent
if $A$, $B$ are in a norm-dense $(\alpha^t)$-invariant $*$-subalgebra of ${\cal A}_\alpha$.
\medskip
	(b) if $A$, $B\in{\cal A}$ there is a continuous function $F_{AB}:\{z\in{\bf C}:0\le{\rm Im}z\le\beta\}\to {\bf C}$, analytic in the strip $\{z\in{\bf C}:0<{\rm Im}z<\beta\}$ and such that 
$$	F_{AB}(t)=\rho(A\alpha^t(B))      $$
$$	F_{AB}(t+i\beta)=\rho(\alpha^t(B)A)      $$
for all $t\in{\bf R}$.}
\medskip
	These conditions also imply that $\rho$ is $(\alpha^t)$-invariant and that 
$$	\sup_z|F_{AB}(z)|\le||A||.||B||      $$
See [1] Definition 5.3.1, Propositions 5.3.3, 5.3.7.
\medskip
	A state $\rho$ satisfying the conditions of the theorem is called a KMS state at value $\beta$ (or at inverse temperature $\beta$).  Note that a KMS state at value $\beta$ for $(\alpha^t)$ is a KMS state at value $\gamma\beta$ for $(\alpha^{\gamma t})$.  Using $\gamma<0$ we may thus also define KMS states at value $\beta<0$ (they correspond to analyticity of $F_{AB}$ in a strip $\{z\in{\bf C}:\beta<{\rm Im}z<0\}$).  By convention a KMS state at value -1 is just called a KMS state.  This absurd terminology is due to a different sign preference of the physicists (in studying the KMS condition) and mathematicians (in studying the modular group).  In any case, since different values of $\beta$ are interchanged by rescaling the time $t$, the normalization $\beta=-1$ is as good as any.
\medskip
	We describe now the relation between the KMS condition and the modular group of the Tomita-Takesaki theory.  Let $\rho$ be a KMS state on the B$^*$-algebra ${\cal A}$ for the automorphism group $(\alpha^t)$, and let $({\cal H}_\rho,\pi_\rho,\Omega_\rho,U_\rho)$ be obtained from these data by the GNS construction.  A normal state $\hat\rho$ and a group of automorphisms $(\hat\alpha^t)$ are defined on the von Neumann algebra $\pi_\rho({\cal A})''$ by 
$$	\hat\rho(M)=(\Omega_\rho,M\Omega_\rho)\qquad,
	\qquad\hat\alpha^t M=U_\rho(t) M U_\rho(t)^{-1}      $$
and we have 
$$	\hat\alpha^t\pi_\rho(A)=\pi_\rho(\alpha^tA)      $$
when $A\in{\cal A}$, $t\in{\bf R}$.  Furthermore it is readily seen that $\hat\rho$ is a KMS state on $\pi_\rho({\cal A})''$ with respect to $(\hat\alpha^t)$.
\medskip
	{\bf Lemma.}
\medskip
	{\sl $\Omega_\rho$ is separating for $\pi_\rho({\cal A})''$.}
\medskip
	Let indeed $A$, $B_1$, $B_2\in\pi_\rho({\cal A})''$.  If $A\Omega_\rho=0$ we have $(\Omega_\rho,(\hat\alpha^tB_2)B_1^*A\Omega_\rho)=0$ hence, by the KMS condition $(\Omega_\rho,B_1^*AB_2\Omega_\rho)=0$ and since $B_1\Omega_\rho$, $B_2\Omega_\rho$ are dense in ${\cal H}_\rho$, this implies $A=0$.\qed 
\medskip
	The vector $\Omega_\rho$ being cyclic and separating for the von Neumann algebra $\pi_\rho({\cal A})''$, a modular group is defined.  Using the modular condition of Section 6 and the KMS condition for $(\hat \alpha^t)$ we find 
$$	(\Delta^{1/2}A\Omega_\rho,\Delta^{1/2}B\Omega_\rho)
	=(\Omega_\rho,BA^*\Omega_\rho)
	=(\Omega_\rho,A^*\hat\alpha^{-i}B\Omega_\rho)
	=(A\Omega_\rho,e^HB\Omega_\rho)      $$
where we have written $U(t)=e^{iHt}$.  Checking domain questions (see [1] Theorem 5.3.10) one obtains that $\Delta=e^H$ as an equality of self-adjoint operators.  Therefore the automorphism group $(\alpha^t)$ of $\pi_\rho({\cal A})''$ coincides with the modular group associated with $\hat\rho$.  In particular, $\rho$ determines uniquely the map $(A,t)\mapsto\pi_\rho(\alpha^tA)=\hat\alpha^t\pi_\rho(A)$ where $A\in{\cal A}$, $t\in{\bf R}$.
\medskip
	{\bf Proposition.}
\medskip
	{\sl If the B$^*$-algebra ${\cal A}$ is simple and $\rho$ is a KMS state on ${\cal A}$ with respect to $(\alpha^t)$, then $(\alpha^t)$ is uniquely determined by $\rho$.}
\medskip
	To check this surprising result, note that $\pi_\rho(\alpha^tA)$ is uniquely determined by $\rho$, and $\pi_\rho$ is injective.\qed
\medskip
	{\bf 8. The set of KMS states.}
\medskip
	Given ${\cal A}$ and $(\alpha^t)$ let $K_\beta$ be the set of KMS states at value $\beta$.  Examples are known where $K_\beta=\emptyset$.  In general $K_\beta$ is convex and $w^*$-compact.  Of particular interest for physical applications is the decomposition of a KMS state into {\it irreducible} KMS states.  We shall only give an informal discussion of this point (see [1] Part 4 for precise definitions and details).
\medskip
	One handles differently the decomposition of a state on a B$^*$-algebra and a normal state on a von Neumann algebra.  Suppose that ${\cal A}$ is a separable B$^*$-algebra, and that $K$ is a convex and $w^*$-compact set of states on ${\cal A}$.  The interesting case is when every $\rho\in K$ has a unique barycentric decomposition into extremal elements of $K$: one then says that $K$ is a {\it simplex}.  Given a von Neumann algebra ${\cal M}$, the interesting case is when an abelian algebra ${\cal C}\subset{\cal M}'$ is given.  Diagonalizing ${\cal C}$ produces a decomposition of a normal state on ${\cal M}$ into states for which the algebra corresponding to ${\cal C}$ is now reduced to multiples of ${\bf 1}$.  
\medskip
	In the case of a KMS state $\rho$, the von Neumann algebra to consider is $\pi_\rho({\cal A})''$, and it turns out that the abelian subalgebra of the commutant which one has to diagonalize is the center $\pi_\rho({\cal A})'\cap\pi_\rho({\cal A})''$.  In other words it turns out that the decomposition of a KMS state into irreducible KMS states is the central decomposition.  A KMS state is thus irreducible, or extremal, if and only if $\pi_\rho({\cal A})'\cap\pi_\rho({\cal A})''=\hbox{multiples of }{\bf 1}$, {\it i.e.}, if $\rho$ is a {\it factor state}.
\medskip
	{\bf Theorem.}
\medskip
	{\sl Let ${\cal A}$ be a B$^*$-algebra with unit element, $(\alpha^t)$ a strongly continuous one-parameter group of homomorphisms and $K_\beta$ the set of KMS states at value $\beta$.  Then
\medskip
	(a) $K_\beta$ is convex, $w^*$-compact, and a simplex,
\medskip
	(b) $\rho$ is an extremal point of $K_\beta$ if and only if it is a factor state.}
\medskip
	For the proof see [1] Theorem 5.3.30.
\medskip
	{\bf 9. Tensor products.{\rm\footnote{*}{See [1] Section 2.7.2.}}.}
\medskip
	If ${\cal H}_1,\ldots,{\cal H}_n$ are Hilbert spaces, and $x_j$, $y_j\in{\cal H}_j$, we write 
$$	(x_1\otimes\cdots\otimes x_n,y_1\otimes\cdots\otimes y_n)
	=\prod_{j=1}^n(x_j,y_j)      $$
This extends to an inner product on the algebraic tensor product of the ${\cal H}_j$ and, after completion, one obtains the {\it Hilbert space tensor product} of ${\cal H}_1,\ldots,{\cal H}_n$ which we denote by ${\cal H}_1\otimes\cdots\otimes{\cal H}_n=\otimes_{j=1}^n{\cal H}_j$.
\medskip
	If ${\cal A}_1,\ldots,{\cal A}_n$ are $C^*$-algebras on ${\cal H}_1,\ldots,{\cal H}_n$, and $A_j\in{\cal A}_j$, the operators $A_1\otimes\cdots\otimes A_n$ on ${\cal H}_1\otimes\cdots\otimes{\cal H}_n$ generate a $C^*$-algebra which we denote by ${\cal A}_1\otimes\cdots\otimes{\cal A}_n=\otimes_{j=1}^n{\cal A}_j$ and call {\it $C^*$-tensor product} of the ${\cal A}_j$.
\medskip
	Let ${\cal B}_1,\ldots,{\cal B}_n$ be $B^*$-algebras and $\pi_j$ a faithful representation of ${\cal B}_j$ in ${\cal H}_j$.  The algebraic tensor product of the ${\cal B}_j$ extends to a $B^*$-algebra by means of the map $\pi_1\otimes\cdots\otimes\pi_n$ into the $C^*$-tensor product $\pi_1({\cal B}_1)\otimes\cdots\otimes\pi_n({\cal B}_n)$.  This extension may be called the {\it $B^*$-tensor product} ${\cal B}_1\otimes\cdots\otimes{\cal B}_n=\otimes_{j=1}^n{\cal B}_j$;  it is unique in the sense that it is (up to isomorphism) independent of the faithful representations $\pi_j$.
\medskip
	If $\sigma_j$ is a state on ${\cal B}_j$, the tensor product of the $\sigma_j$ extends to a unique state $\sigma=\sigma_1\otimes\cdots\otimes\sigma_n$ on ${\cal B}_1\otimes\cdots\otimes{\cal B}_n$, and the cyclic representation $({\cal H}_\sigma,\pi_\sigma,\Omega_\sigma)$ associated with $\sigma$ is in a natural manner the tensor product of the $({\cal H}_{\sigma_j},\pi_{\sigma_j},\Omega_{\sigma_j})$.
\medskip
	If ${\cal M}_1,\ldots,{\cal M}_n$ are von Neumann algebras on ${\cal H}_1,\ldots,{\cal H}_n$, the weak closure of the algebraic tensor product of ${\cal M}_1,\ldots,{\cal M}_n$ acting on ${\cal H}_1\otimes\ldots\otimes{\cal H}_n$ will be called their {\it von Neumann tensor product} ${\cal M}_1\otimes\ldots\otimes{\cal M}_n$.
\medskip
	{\bf Theorem} (commutation for tensor products).
\medskip
	{\sl Let ${\cal M}_1$, ${\cal M}_2$ be von Neumann algebras on ${\cal H}_1$, ${\cal H}_2$ with commutants ${\cal M}'_1$, ${\cal M}'_2$, then the commutant $({\cal M}_1\otimes{\cal M}_2)'$ of ${\cal M}_1\otimes{\cal M}_2$ on ${\cal H}_1\otimes{\cal H}_2$ is the von Neumann tensor product ${\cal M}'_1\otimes{\cal M}'_2$.}
\medskip
	Obviously, this extends to $n$ factors.  The above commutation theorem was first proved by Takesaki using the Tomita-Takesaki theory.  See [3] Theorem 2.8.1 for another proof.
\medskip
	{\bf Example} (factor states).
\medskip
	A state $\sigma$ on a $B^*$-algebra ${\cal A}$ is called a factor state if $\pi_\sigma({\cal A})''$ is a factor.  A tensor product $\sigma=\sigma_1\otimes\cdots\otimes\sigma_n$ of factor states is again a factor state.  Indeed $\pi_\sigma({\cal A}_1\otimes\cdots\otimes{\cal A}_n)''\cup\pi_\sigma({\cal A}_1\otimes\cdots\otimes{\cal A}_n)'=\pi_{\sigma_1}({\cal A}_1)''\otimes\cdots\otimes\pi_{\sigma_n}({\cal A}_n)''\cup\pi_{\sigma_1}({\cal A}_1)'\otimes\cdots\otimes\pi_{\sigma_n}({\cal A}_n)'$
which is irreducible.
\medskip
	{\bf References.}

[1] O. Bratteli and D.W. Robinson.  {\it Operator algebras and quantum statistical mechanics I, II.}  Springer, New York, 1979-1981.  [There is a 2-nd ed. (1997) of vol. II].

[2] M. Reed and B. Simon.  {\it Methods of modern mathematical physics I, II, III, IV.}  Academic Press, New York, 1972-1975-1979-1978.

[3] Sh. Sakai.  {\it C$^*$-algebras and W$^*$-algebras.}  Springer, Berlin, 1971.
\vfill\eject
\null\bigskip\bigskip
\centerline{Chapter 2. QUANTUM SPIN SYSTEMS.}
\bigskip\bigskip
	{\bf 1. Quasilocal structure.}
\medskip
	In what follows, $L$ will be a countably infinite set.  We view $L$ as a ``lattice'' ($L={\bf Z}^\nu$ is a standard example), and the points $x\in L$ are sites at which quantum spins are located.  For each $x\in L$ let a finite dimensional complex hilbert space ${\cal H}_x$ be given.  For finite $\Lambda\subset L$ we define 
$$	{\cal H}_\Lambda=\otimes_{x\in\Lambda}{\cal H}_x      $$
and let ${\cal A}_\Lambda$ be the algebra ${\cal L}({\cal H}_\Lambda)$ of bounded operators on ${\cal H}_\Lambda$.  If $\Lambda_1\subset\Lambda_2$, there is an isomorphism ${\cal A}_{\Lambda_1}\mapsto{\cal A}_{\Lambda_1}\otimes{\bf 1}_{\Lambda_2\backslash\Lambda_1}\subset{\cal A}_{\Lambda_2}$, which we can use to identify the B$^*$-algebra ${\cal A}_{\Lambda_1}$ to a subalgebra of ${\cal A}_{\Lambda_2}$.  In this manner $\cup_\Lambda{\cal A}_\Lambda$ is a normed $*$-algebra, and its norm completion ${\cal A}$ is a B$^*$-algebra.
\medskip
	The following properties hold:

\noindent(1)$\qquad\Lambda_1\subset\Lambda_2\qquad\Rightarrow\qquad{\cal A}_{\Lambda_1}\subset{\cal A}_{\Lambda_2}$

\noindent(2)$\qquad\cup_\Lambda{\cal A}_\Lambda\hbox{ is dense in }{\cal A}$

\noindent(3)$\qquad{\cal A}\hbox{ and all }{\cal A}_\Lambda\hbox{ have a common unit }{\bf 1}$

\noindent(4)$\qquad\Lambda_1\cap\Lambda_2=\emptyset\qquad\Rightarrow\qquad[{\cal A}_{\Lambda_1},{\cal A}_{\Lambda_2}]=0$

\noindent(5)$\qquad{\cal A}\hbox{ is simple}$

Properties (1) -- (4) express that ${\cal A}$, equipped with the net $({\cal A}_\Lambda)$ of B$^*$-algebras is a {\it quasilocal algebra} (the ${\cal A}_\Lambda$ are the {\it local algebras}).  Property (5) means that if $\pi$ is a nonzero morphism ${\cal A}\to{\cal B}$, its kernel is 0 [this is a consequence of the fact that the ${\cal A}_\Lambda$ are simple; see [1] Section 2.6.3].
\medskip
	Suppose now that $L={\bf Z}^\nu$ and that all ${\cal H}_x$ are copies of ${\cal H}_0$ so that for all $a\in{\bf Z}^\nu$ there are canonical unitary maps $U_a:{\cal H}_\Lambda\to{\cal H}_{\Lambda+a}$ and isomorphisms $\tau_a:A\mapsto U_aAU_a^{-1}$ of ${\cal A}_\Lambda$ to ${\cal A}_{\Lambda+a}$, extending to automorphisms $\tau_a:{\cal A}\to{\cal A}$ with the group properties $\tau_0=$ identity, $\tau_a\tau_b=\tau_{a+b}$.  From (2), (4) one gets readily 
$$	\lim_{a\to\infty}||[A,\tau_aB]||=0
	\qquad{\rm if}\qquad A,B\in{\cal A}      $$
This property is known as {\it asymptotic abelianness} (in the norm sense).
\medskip
	{\bf 2. Time evolution.}
\medskip
	A function $\Phi$ from the finite subsets of $L$ to ${\cal A}$ such that $\Phi(X)=\Phi(X)^*\in{\cal A}_X$ is called an {\it interaction}.  For finite $\Lambda$ we define the associated {\it Hamiltonian}
$$	H_\Phi(\Lambda)=\sum_{X\subset\Lambda}\Phi(X)      $$
$H_\Phi(\Lambda)$ is thus a self-adjoint element of ${\cal A}_\Lambda$.
\medskip
	If $L={\bf Z}^\nu$ we say that the interaction $\Phi$ is {\it translationally invariant} when
$$	\Phi(X+a)=\tau_a\Phi(X)      $$
(for all finite $X\subset L$ and all $a\in{\bf Z}^\nu$).
\medskip
	Under various boundedness conditions, an interaction $\Phi$ defines a {\it time evolution}, {\it i.e.}, a one-parameter group of $*$-automorphisms of ${\cal A}$.  Here we shall assume that 
$$	||\Phi||_\lambda=\sum_{n\ge0}e^{n\lambda}\sup_{x\in L}
	\sum_{X\ni x:{\rm card}X=n+1}||\Phi(X)||<+\infty      $$
for some $\lambda>0$ (such interactions form a Banach space with norm $||\cdot||_\lambda$).
\medskip
	Before defining the time evolution associated with $\Phi$, we introduce what should be its infinitesimal operator.  Let $D(\delta)=\cup_\Lambda{\cal A}_\Lambda$ and let $\delta:D(\delta)\to{\cal A}$ be such that
$$	\delta(A)=i\sum_{X:X\cap\Lambda\ne\emptyset}[\Phi(X),A]
	\qquad{\rm if}\qquad A\in{\cal A}_\Lambda      $$
We have 
$$	||\delta(A)||\le\sum_{x\in\Lambda}\sum_{X\ni x}2||\Phi(X)||.||A||
	\le2|\Lambda|.||A||\sup_{x\in L}||\Phi(X)||
	\le2|\Lambda|.||A||.||\Phi(X)||_\lambda      $$
so that $\delta$ is well defined on the dense domain $D(\delta)$, and
$$	\delta(A^*)=\delta(A)^*      $$
$$	\delta(AB)=\delta(A)B+A\delta(B)      $$
{\it i.e.}, $\delta$ is a symmetric derivation.
\medskip
	We also have, if $A\in{\cal A}_\Lambda$,
$$	\delta^m(A)=i^m{\textstyle\sum^*_{X_m}\cdots\sum^*_{X_1}}
	[\Phi(X_m),[\cdots[\Phi(X_1),A]\ldots]]      $$
where $\sum^*_{X_j}$ extends over those $X_j$ such that
$$	X_j\cap S_{j-1}\ne\emptyset      $$
and
$$	S_0=\Lambda\qquad,\qquad S_j=X_j\cup S_{j-1}\hbox{ for $i\ge1$}      $$
We may write 
$$	{\textstyle\sum_{X_j}^*=\sum_{n_j=0}^\infty\sum_{X_j}^{**}}      $$
where $\sum_{X_j}^{**}$ extends over those $X_j$ such that 
$$	X_j\cap S_{j-1}\ne0\qquad,\qquad{\rm card}X_j=n_j+1      $$
In particular,
$$	{\rm card}S_j
	\le{\rm card}\Lambda+{\rm card}X_1+\ldots+{\rm card}X_j-j
	={\rm card}\Lambda+n_1+\ldots+n_j      $$
so that
$$	||\delta^m(A)||\le2^m||A||\sum_{n_1,\ldots,n_m\ge0}
	\prod_{j=1}^m({\rm card}\Lambda+n_1+\ldots+n_{j-1})\prod_{j=1}^m
	\sup_{x\in L}\sum_{X_j\ni x,{\rm card}X_j=n_j+1}||\Phi(X_j)||      $$
We have 
$$	\prod_{j=1}^m({\rm card}\Lambda+n_1+\ldots+n_{j-1})
	\le({\rm card}\Lambda+n_1+\ldots+n_m)^m      $$
$$   \le m!\lambda^{-m}\exp[\lambda({\rm card}\Lambda+n_1+\ldots+n_m)]   $$
and therefore
$$	||\delta^m(A)||\le||A||e^{\lambda{\rm card}\Lambda}m!
	(2\lambda^{-1}||\Phi||_\lambda)^m      $$
We have proved the following
\medskip
	{\bf Lemma.}
\medskip
	{\sl If $A\in D(\delta)$, the series
$$	\sum_{m=0}^\infty{t^m\over m!}||\delta^m(A)||
	\le||A||e^{\lambda{\rm card}\Lambda}\sum_{m=0}^\infty 
	t^m(2\lambda^{-1}||\Phi||_\lambda)^m      $$
converges when $|t|<\lambda/2||\Phi||_\lambda$.}
\medskip
	The elements in the domain $D(\lambda)=\cup_\Lambda{\cal A}_\Lambda$ are thus {\it analytic vectors} for $\delta$.
\medskip
	{\bf Theorem} (existence of time evolution).
\medskip
	{\sl If $||\Phi||_\lambda<\infty$, there is a strongly continuous}\footnote{*}{This means that for each $A$, $t\mapsto\alpha^tA$ is continuous ${\bf R}\to{\cal A}$ (with the norm topology of ${\cal A}$).} {\sl one-parameter group $(\alpha^t)$ of automorphisms of ${\cal A}$ such that
$$	\alpha^tA=\sum_{m=0}^\infty{t^m\over m!}\delta^mA\eqno{(1)}      $$
if $A\in\cup_\Lambda{\cal A}_\Lambda$ and $|t|<\lambda/2||\Phi||_\lambda$.  If we define
$$	\alpha_\Lambda^tA=e^{itH_\Phi(\Lambda)}Ae^{-itH_\Phi(\Lambda)}      $$
we have 
$$	\lim_{\Lambda\to\infty}||\alpha^t(A)-\alpha_\Lambda^t(A)||
	=0\eqno{(2)}      $$
for all $A\in{\cal A}$, uniformly for $t$ in compacts.}
\medskip
	Under the conditions
$$	A\in{\cal A}_\Lambda\qquad,\qquad
	|t|<\lambda/2||\Phi||_\lambda\eqno{(3)}      $$
the expression of $\alpha_\Lambda^t(A)$ in powers of $t$ tends term by term to (1) when $\Lambda\to\infty$.  Because of uniform bounds, we have thus (2) when (3) holds.  In particular, in
$$	\{t:|t|<\lambda/2||\Phi||_\Lambda\}\eqno{(4)}      $$
we can extend $\alpha^t$ by continuity to an automorphism of ${\cal A}$.  Furthermore $t\mapsto\alpha^tA$ is continuous in (4) and 
$$	\alpha^0={\rm identity}\qquad,\qquad
	\alpha^s\alpha^t=\alpha^{s+t}      $$
for $s$, $t$, $s+t$ in (4), and this permits an extension of $\alpha^t$ to all $t\in{\bf R}$ so that the group property $\alpha^s\alpha^t=\alpha^{s+t}$ is satisfied.  The uniformity of (2) is readily checked.\qed
\medskip
	{\bf Remark.}
\medskip
	The above proof follows [2] Section 7.6.  In [1] the corresponding Theorem 6.2.4 results from a more general formalism, and it is shown that the infinitesimal generator of $(\alpha^t)$ is the closure $\bar\delta$ of $\delta$.  [The infinitesimal generator $S$ of $(\alpha^t)$ is defined by 
$$	SA=\lim_{t\to0}{\alpha^tA-A\over t}      $$
whenever the limit exists in the norm topology of ${\cal A}$ (see [1] Corollary 3.1.8).  That $\delta$ is closable follows from [1] Proposition 3.2.22, Lemma 3.1.14].
\medskip
	{\bf 3. Digression}\footnote{*}{See [3]}{\bf : the algebras ${\cal A}_\lambda$.}
\medskip
	For finite $\Lambda\subset L$, a map $\pi_\Lambda:\cup_X{\cal A}_X\to{\cal A}_\Lambda$ is defined by
$$	\pi_\Lambda A=\lim_{Y\to L\backslash\Lambda}\,
	{{\rm tr}_{{\cal H}_Y}A\over{\rm dim}{\cal H}_Y}      $$
If the $\phi_i$ form an orthonormal basis of ${\cal H}_Y$, and $\psi',\psi''\in
{\cal H}_\Lambda$ we have 
$$	(\psi',{{\rm tr}_{{\cal H}_Y}A\over{\rm dim}{\cal H}_Y}\psi'')
={1\over{\rm dim}{\cal H}_Y}\sum_i(\phi_i\otimes\psi',A\phi_i\otimes\psi'') $$
hence $||\pi_\Lambda A||\le||A||$.  The properties of the following lemma are then readily checked.
\medskip
	{\bf Lemma}
\medskip
	{\sl The map $\pi_\Lambda$ extends to a unique linear norm-reducing map ${\cal A}\to{\cal A}_\Lambda$.  Furthermore
$$	\pi_\Lambda A=A\qquad{\rm if}\qquad A\in{\cal A}_\Lambda      $$
$$	\pi_\Lambda A^*=(\pi_\Lambda A)^*      $$
$$	\pi_\Lambda\pi_{\Lambda'}=\pi_{\Lambda'}\pi_\Lambda      $$}
\indent
	Choose now some $\lambda>0$.  For $A\in{\cal A}_\Lambda$, define
$$	||A||_\lambda=\inf\{\sum_{X\subset\Lambda}
	||A_X||e^{\lambda\,{\rm card}X}:\sum_X A_X=A\}      $$
By compactness we may replace the inf by min.  If $\Lambda$ is replaced by a larger set $\Lambda'$, and $\sum_Y A_Y=A$ with $Y\subset\Lambda'$, we have
$$	\sum_{Y\subset\Lambda'}||A_Y||e^{\lambda\,{\rm card}Y}
	\ge\sum_Y||\pi_\Lambda A_Y||e^{\lambda\,{\rm card}(Y\cap\Lambda)}    $$
with $\sum_Y\pi_\Lambda A_Y=\pi_\Lambda A=A$.  Therefore $||A||_\lambda$ does not depend on the choice of $\Lambda$ provided $A\in{\cal A}_\Lambda$.  We have thus a norm $||.||_\lambda$ on $\cup_X{\cal A}_X$, and we may define the Banach space ${\cal A}_\lambda$ by completion.
\medskip
	{\bf Proposition.}
\medskip
	{\sl The inclusion map $\cup_X{\cal A}_X\to{\cal A}$ extends to a norm-reducing map $\omega:{\cal A}_\lambda\to{\cal A}$ and $\omega$ is injective.}
\medskip
	$\omega$ is norm-reducing because $||A||\le||A||_\lambda$ for $A\in\cup_X{\cal A}_X$.
\medskip
	Note now that $\pi_\Lambda:\cup_X{\cal A}_X\to{\cal A}_\Lambda$ reduces the $||.||_\lambda$-norm and extends thus to a linear norm-reducing map ${\cal A}_\lambda\to{\cal A}_{\Lambda\lambda}$ where ${\cal A}_{\Lambda\lambda}$ is ${\cal A}_\Lambda$ equipped with the $||.||_\lambda$-norm.  Assume that $A\in{\cal A}_\lambda$ with $||A||_\lambda=a>0$.  We may choose $\Lambda$ and $B\in{\cal A}_\Lambda$ such that $||A-B||_\lambda<a/3$, hence $||B||_\lambda>2a/3$.  Now $\omega A=0$ would imply $\pi_\Lambda A=0$ hence 
$$	{2a\over3}<||B||_\lambda=||\pi_\Lambda(B-A)||_\lambda
	\le||A-B||_\lambda<{a\over3}      $$
Therefore $\omega$ must be injective.\qed
\medskip
	{\bf Corollary.}
\medskip
	{\sl ${\cal A}_\lambda$ is identified by $\omega$ to a dense $*$-subalgebra of ${\cal A}$; ${\cal A}_\lambda$ is then a Banach algebra with respect to the norm $||.||_\lambda$.  Taking $\lambda=0$ we may define ${\cal A}_0={\cal A}$.  With this definition, if $\lambda<\mu$ we have ${\cal A}_\lambda\supset{\cal A}_\mu$, and the map ${\cal A}_\mu\to{\cal A}_\lambda$ is norm-reducing.}
\medskip
	If $A,B\in{\cal A}_\Lambda$ we may choose $A_X,B_X\in{\cal A}_X$ such that $A=\sum_{X\subset\Lambda}A_X$, $B=\sum_{X\subset\Lambda}B_X$, and 
$$  ||A||_\lambda=\sum_{X\subset\Lambda}||A_X||e^{\lambda\,{\rm card}X}\qquad,
\qquad||B||_\lambda=\sum_{X\subset\Lambda}||B_X||e^{\lambda\,{\rm card}X}  $$
Thus
$$ ||AB||_\lambda\le\sum_X\sum_Y||A_XA_Y||e^{\lambda\,{\rm card}(X\cup Y)} $$
$$	\le\sum_X\sum_Y||A_X||.||A_Y||e^{\lambda({\rm card}X+{\rm card}Y)}
	=||A||_\lambda||B||_\lambda      $$
Therefore if $A$, $B$ tend to limits $A_\infty$, $B_\infty$ in ${\cal A}_\lambda$, $AB$ tends in ${\cal A}_\lambda$ to $A_\infty B_\infty$ and $||A_\infty B_\infty||_\lambda$ $\le||A_\infty||_\lambda||B_\infty||_\lambda$.  The rest is clear.\qed
\medskip
	{\bf Proposition.}
\medskip
	{\sl Suppose that $\lambda>\mu\ge0$ and $||\Phi||_\lambda<\infty$.  Then
$$	||\delta(A)||_\mu
	\le2(\lambda-\mu)^{-1}||A||_\lambda||\Phi||_\mu      $$
$$	||\delta^m(A)||_\mu
	\le||A||_\lambda m!(2(\lambda-\mu)^{-1}||\Phi||_\lambda)^m      $$
In particular $\alpha^tA\in{\cal A}_\mu$ if $|t|<(\lambda-\mu)/2||\Phi||_\lambda$.}
\medskip
	The proof follows basically the estimates in Section 2.\qed
\medskip
	{\bf 4. Perturbation of the time evolution.}\footnote{*}{See [1] Section 5.4.1.}
\medskip
	A strongly continuous one-parameter group $(\alpha^t)$ of $*$-automorphisms of a B$^*$-algebra ${\cal A}$ is entirely determined by its infinitesimal generator $S$ (which is a densely defined derivation).  If $P=P^*\in{\cal A}$, the derivation
$$	A\mapsto SA+i[P,A]      $$
is the generator of a new strongly continuous one-parameter group $(\alpha_P^t)$ of automorphisms of ${\cal A}$.  We shall not justify this assertion, but note that if ${\cal A}$ is a C$^*$-algebra on ${\cal H}$, and if $(\alpha^t)$ is unitarily implemented, {\it i.e.}, $U(t)=e^{iHt}$ is a one-parameter group of unitary operators on ${\cal H}$ such that 
$$	\alpha^tA=U(t)AU(-t)      $$
then
$$	\alpha_P^tA=U_P(t)AU_P(-t)\eqno{(5)}      $$
with $U_P(t)=e^{i(H+P)t}$.  [It is however not clear from (5) that $t\mapsto\alpha_P^tA$ is norm continuous].
\medskip
	The group $(\alpha^t_P)$ is determined by the integral equation
$$	\alpha^t_PA
=\alpha^tA+i\int_0^td\tau\,\alpha^\tau[P,\alpha^{t-\tau}_PA]      $$
with solution given by
$$	\alpha^t_PA=\alpha^tA+\sum_{n=1}^\infty i^n
	\int_0^tdt_1\int_0^{t_1}dt_2\cdots\int_0^{t_{n-1}}dt_n\,
[\alpha^{t_n}P,[\alpha^{t_{n-1}}P,[\cdots[\alpha^{t_1}P,\alpha^tA]\cdots]]   $$
or
$$	\alpha_P^tA=\Gamma_t(\alpha^tA)\Gamma_{-t}      $$
where $(\Gamma_t)$ is a one-parameter family of unitary elements of ${\cal A}$ such that\footnote{*}{The family $(\Gamma_t)$ satisfies the cocycle condition
$\Gamma_{t+s}=\Gamma_t\,(\alpha^t\,\Gamma_s)$.}
$$	\Gamma_t={\bf 1}+\sum_{n=1}^\infty i^n
	\int_0^tdt_1\int_0^{t_1}dt_2\cdots\int_0^{t_{n-1}}dt_n\,
	(\alpha^{t_n}P)\cdots(\alpha^{t_1}P)      $$
\indent
	We exhibit now a one-to-one correspondence $\rho\to\rho_P$ between the KMS states at value $\beta>0$ for $(\alpha^t)$ and for the perturbed evolution $(\alpha_P^t)$.  If $({\cal H}_\rho,\pi_\rho,\Omega_\rho)$ is the cyclic representation associated with $\rho$ we write 
$$	\Omega_P=\Omega_\rho+\sum_{n=1}^\infty (-1)^n
	\int_0^{\beta/2}ds_1\int_0^{\beta/2}ds_2\cdots\int_0^{\beta/2}ds_n\,
	\pi_\rho((\alpha^{is_n}P)\cdots(\alpha^{is_1}P))\Omega_\rho      $$
Then the state $\rho_P$ is defined by
$$	\rho_P(A)={(\Omega_P,\pi_\rho(A)\Omega_P)\over(\Omega_P,\Omega_P)}   $$
One can show ([1] Corollary 5.4.5) that $\rho\mapsto\rho_P$ is an isomorphism of the set of KMS states at value $\beta$ for $(\alpha^t)$ and the set of KMS states at value $\beta$ for $(\alpha_P^t)$.  This map sends extremal KMS states to extremal KMS states, but is not affine.  We note also the formula
$$	\rho_P(A)=\rho(A)+\sum_{n=1}^\infty\int_{-1\le s_1\le\cdots\le s_n\le0}
ds_1\cdots ds_n\,\rho(A,(\alpha^{is_n}P),\cdots,(\alpha^{is_1}P))^T      $$
valid if $||P||<1/2$, and where $\rho(A_0,\ldots,A_n)^T$ denotes a {\it truncated expectation value}.
\medskip
	{\bf 5. M\o ller morphisms.}
\medskip
	Under a suitable asymptotic abelianness condition, an affine relation between KMS states for $(\alpha^t)$ and $(\alpha_P^t)$ will now be obtained as the adjoint of an endomorphism $\gamma_\pm$ of the algebra ${\cal A}$.
\medskip
	We say that $(\alpha^t)$ is $L^1({\cal A}_0)$-{\it asymptotically abelian} if
$$	\int_{-\infty}^\infty dt\,||[A,\alpha^tB]||<\infty      $$
for all $A$, $B$ in the norm-dense $*$-subalgebra ${\cal A}_0$ of ${\cal A}$.  This implies in particular that $(\alpha^t)$ is asymptotically abelian in the norm sense:
$$	\lim_{|t|\to\infty}||[A,\alpha^tB]||=0
	\hbox{ for all }A,B\in{\cal A}      $$
\indent
	{\bf Proposition.}
\medskip
	{\sl If $(\alpha^t)$ is $L^1({\cal A}_0)$-asymptotically abelian, the limits 
$$	\gamma_\pm A=\lim_{t\to\pm\infty}\alpha_P^{-t}\alpha^tA      $$
exist in norm for all $A\in{\cal A}$ and $P=P^*\in{\cal A}_0$.  Th maps $\gamma_\pm$ are called M\o ller morphisms: they are norm-preserving $*$-morphisms ${\cal A}\to{\cal A}$ which satisfy the intertwining relations
$$	\gamma_\pm\alpha^t=\alpha_P^t\gamma_\pm      $$
If ${\cal A}$ has a unit element ${\bf 1}$, the adjoints $\gamma_\pm^*$ map the set $E$ of states on ${\cal A}$ into itself, and 
\medskip
	(1) $(\alpha_P^t)$-invariant (resp. ergodic) states are mapped to $(\alpha^t)$-invariant (resp. ergodic) states,
\medskip
	(2) KMS states (resp. extremal KMS states) for $(\alpha_P^t)$ are mapped to KMS states (resp. extremal KMS states) for $(\alpha^t)$, furthermore $\gamma_+^*$ and $\gamma_-^*$ coincide on the $(\alpha_P^t)$-KMS states.}
\medskip
	See [1] Proposition 5.4.10.
\medskip
	{\bf 6. Gibbs states.}
\medskip
	Given an interaction $\Phi$, and finite $\Lambda\subset L$, the {\it local Gibbs state} $\rho_\Lambda^\Phi$ is a state on ${\cal A}_\Lambda$ defined by 
$$	\rho_\Lambda^\Phi(A)={Tr_{{\cal H}_\Lambda}(e^{-H_\Phi(\Lambda)}A)
	\over Tr_{{\cal H}_\Lambda}e^{-H_\Phi(\Lambda)}}      $$
Replacing $\Phi$ by $\beta\Phi$ we obtain the usual definition of the local Gibbs state $\rho_\Lambda^{\beta\Phi}$ at inverse temperature $\beta$.  When $\Lambda\to\infty$ in the sense that each finite subset of $L$ is eventually contained in $\Lambda$) $\rho_\Lambda^{\beta\Phi}$ has limit points $\rho^{\beta\Phi}$
(in the sense that $\rho_\Lambda^{\beta\Phi}(A)\to\rho^{\beta\Phi}(A)$ for $A\in\cup_\Lambda{\cal A}_\Lambda$).  Under our condition $||\Phi||_\lambda<\infty$, it follows that every limit point $\rho^{\beta\Phi}$ is a KMS point (at value $\beta$) on ${\cal A}$ (see Proposition [1] 6.2.15).  In particular, in the situation that we consider, there are always KMS states, sometimes a single one, sometimes many.
\medskip
	Given the interaction $\Phi$ (satisfying $||\Phi||_\lambda<\infty$) and finite $\Lambda\subset L$ we define $W_\Phi(\Lambda)\in{\cal A}$ by 
$$	W_\Phi(\Lambda)=\textstyle{\sum_{X\subset\Lambda}^*}\Phi(X)      $$
where $\sum^*$ extends over those $X$ not contained in $\Lambda$ or in its complement $L\backslash\Lambda$.  We may thus interpret $W_\Phi(\Lambda)$ as the energy of interaction between $\Lambda$ and $L\backslash\Lambda$.  Formally, we have
$$   H_\Phi(L)=H_\Phi(\Lambda)+W_\Phi(\Lambda)+H_\Phi(L\backslash\Lambda)   $$
Let a Gibbs state $\rho$ on ${\cal A}$ be formally defined by 
$$	\rho(A)={Tr(e^{-H_\Phi(L)}A)\over Tr e^{-H_\Phi(L)}}      $$
If we remove the interaction between $\Lambda$ and $L\backslash\Lambda$, we see that $\rho$ should factorize as \break $\rho_\Lambda^\Phi\otimes(\hbox{state on }{\cal A}_{L\backslash\Lambda})$.  This leads to the following definition.
\medskip
	A state $\rho$ on ${\cal A}$ is a {\it Gibbs state} for $\beta\Phi$ if 
\medskip
	(1) $\rho$ is faithful, {\it i.e.}, $\Omega_\rho$ is separating for $\pi_\rho({\cal A})''$,
\medskip
	(2) for all finite $\Lambda\subset L$, if $P=\beta W_\Phi(\Lambda)$, then
$$	\rho_P=\rho_\Lambda^\Phi\otimes\tilde\rho      $$
where $\tilde\rho$ is a state over ${\cal A}_{L\backslash\Lambda})$. 
\medskip
	{\bf Theorem} (equivalence of Gibbs and KMS states).
\medskip
	{\sl Let the interaction $\Phi$ satisfy $||\Phi||_\lambda<\infty$ for some $\lambda>0$ and $(\alpha^t)$ be the corresponding time evolution on ${\cal A}$.  The following are equivalent:
\medskip
	(a) $\rho$ is a Gibbs state with respect to $\beta\Phi$,
\medskip
	(b) $\rho$ is a KMS state with respect to $(\alpha^t)$ at value $\beta$.}
\medskip
	See [1] Corollary 6.2.19.
\medskip
	It is also possible to characterize the KMS or Gibbs states by a ``maximum entropy principle'' (see [1] Section 6.2.3).  If $L={\bf Z}^\nu$, and one restricts to translationally invariant states one has a theory very analogous to the theory of classical lattice spin systems (see [2]), with equivalence between equilibrium states (satisfying a variational principle) and invariant Gibbs states (see [1] Sections 6.2.4, 6.2.5, 6.2.6).  The ``Notes and Remarks'' on Sections 6.1 and 6.2 at the end of [1] vol II give some perspective on who did what, and indicate in particular the important role of H. Araki.
\medskip
	{\bf References.}

[1] O. Bratteli and D.W. Robinson.  {\it Operator algebras and quantum statistical mechanics I, II.}  Springer, New York, 1979-1981.  [There is a 2-nd ed. (1997) of vol. II].

[2] D. Ruelle.  {\it Statistical mechanics.  Rigorous results.}  Benjamin, New York, 1969.

[3] D.Ruelle.  ``Entropy production in quantum spin systems.''  Preprint.
\vfill\eject
\null\bigskip\bigskip
\centerline{Chapter 3. NONEQUILIBRIUM.}
\bigskip\bigskip
	{\bf 1. Physical model.}\footnote{*}{Sections 1-5 of the present Chapter follow Ruelle [4].}
\medskip
	In this chapter we specialize the setup of Chapter 2.  We first recall the latter briefly.  A countably infinite set $L$ is given; the points of $L$ are interpreted as {\it sites} at which quantum spins are located.  For each $x\in L$, a finite dimensional complex Hilbert space ${\cal H}_x$ is given (describing the spin states at $x$).  For finite $\Lambda\subset L$ we define
$$	{\cal H}_\Lambda=\otimes_{x\in\Lambda}{\cal H}_x      $$
and let ${\cal A}_\Lambda$ be the algebra ${\cal L}({\cal H}_\Lambda)$ of bounded operators on ${\cal H}_\Lambda$.  If $\Lambda_1\subset\Lambda_2$ we identify ${\cal A}_{\Lambda_1}$ to a subalgebra of ${\cal A}_{\Lambda_2}$ by the isomorphism $A\to A\otimes{\bf 1}_{{\Lambda_2}\backslash{\Lambda_1}}$ and let the B$^*$-algebra ${\cal A}$ be the completion of $\cup_\Lambda{\cal A}_\Lambda$.  The algebra ${\cal A}$, equipped with the net $({\cal A}_\Lambda)$ of {\it local algebras}, is a {\it quasilocal algebra}, and the properties (1)--(5) of Chapter 2, Section 1 hold.
\medskip
	We write now $L$ as a finite disjoint union 
$$	L=S+R_1+R_2+\ldots      $$
where $R$ is finite and the $R_a$ (for $a>0$) are infinite.  The physical meaning of the decomposition is as follows: $S$ is a ``small'' system connected to different ``large'' reservoirs $R_a$ ($a=1,2,\ldots$).  We define the quasilocal algebras ${\cal A}_a$ as the norm closures of 
$$	\cup_{X\subset R_a}{\cal A}_X      $$
for $a>0$.  It is convenient to write also $S=R_0$ and ${\cal A}_S={\cal A}_0$.
\medskip
	{\bf 2. Assumptions.}
\medskip
	We assume that an interaction $\Phi:X\mapsto\Phi(X)$ is given such that $\Phi(X)$ is a self-adjoint element of ${\cal A}_X$ for every finite $X\subset L$.  Also, for each reservoir we prescribe an inverse temperature $\beta_a>0$ and a state $\sigma_a$ on ${\cal A}_a$.  These data should satisfy the conditions (A1), (A2), (A3) given below.
\medskip
	(A1) {\sl The interaction $\Phi$ satisfies
$$	||\Phi||_\lambda=\sum_{n\ge0}e^{n\lambda}\sup_{x\in L}
	\sum_{X\ni x:{\rm card}X=n+1}||\Phi(X)||<+\infty      $$
for some $\lambda>0$.}
\medskip
	[This condition permits the definition of a time evolution $(\alpha^t)$ as discussed in Chapter 2].
\medskip
	(A2) {\sl $\Phi(X)=0$ if $X\cap S=\emptyset$, $X\cap R_a\ne\emptyset$, $X\cap R_b\ne\emptyset$ for different $a,b>0$.}
\medskip
	[Note that the description of the interaction $\Phi$ is somewhat ambiguous because anything ascribed to $\Phi(X)$ might be ascribed to $\Phi(Y)$ for some $Y\supset X$.  Condition (A2) means that, in our accounting, if a part of our interaction connects two different reservoirs, it must also involve the small system $S$.  In other words, the reservoirs do not interact directly].
\medskip
	(A3) {\sl If $a>0$, let $\Phi_a$ be the restriction of the interaction $\Phi$ to subsets of $R_a$ and write 
$$	H_{a\Lambda}=\sum_{X\subset R_a\cap\Lambda}\Phi_a(X)
	=H_{R_a\cap\Lambda}      $$
Let also the interactions $\Psi_{(\Lambda)}$ be given such that 
$$	||\Psi_{(\Lambda)}||_\lambda\le K<\infty\eqno{(1)}      $$
and write
$$	B_{a\Lambda}=\sum_{X\subset R_a\cap\Lambda}\Psi_{(\Lambda)}(X)      $$
We assume that, for a suitable sequence $\Lambda\to L$,
$$	\lim_{\Lambda\to L}{{\rm Tr}_{{\cal H}_{R_a\cap\Lambda}}
	(e^{-\beta_a(H_{a\Lambda}+B_{a\Lambda})}A)\over
	{\rm Tr}_{{\cal H}_{R_a\cap\Lambda}}
	e^{-\beta_a(H_{a\Lambda}+B_{a\Lambda})}}=\sigma_a(A)      $$
if $A\in{\cal A}_a$: this defines a state $\sigma_a$ on ${\cal A}_a$, depending on the choice of $(\Psi_{(\Lambda)})$ and the sequence $\Lambda\to L$.  Furthermore we assume that for each finite $X$ there is $\Lambda_X$ such that $\Psi_{(\Lambda)}(Y)=0$ if $\Lambda\supset\Lambda_X$ and $Y\subset X$; therefore
$$	||[B_{a\Lambda},A]||=0\eqno{(2)}      $$
if $\Lambda\supset\Lambda_X$ and $A\in{\cal A}_X$.}
\medskip
	[A possible choice is thus $\Psi_{(\Lambda)}=0$ for all $\Lambda$.  Using (3) below, it is readily verified that $\sigma_a$ is a $\beta_a$-KMS state (see Chapter 1, Section 7) for the one-parameter group $(\breve\alpha_a^t)$ of automorphisms of ${\cal A}_a$ corresponding to the interaction $\Phi_a$.  It is not known which of the $\beta_a$-KMS states can be obtained in this manner].
\medskip
	Note that the assumptions (A1), (A2), (A3) can be explicitly verified in specific cases.  
\medskip
	{\bf 3. Some technical consequences.}
\medskip
	From (A3) we obtain the following result.
\medskip
	{\bf Lemma.}
\medskip{\sl
$$	\lim_{\Lambda\to L}||e^{it(H_{a\Lambda}+B_{a\Lambda})}A
e^{-it(H_{a\Lambda}+B_{a\Lambda})}-\breve\alpha_a^tA||=0\eqno{(3)}      $$
for $a>0$, and
$$	\lim_{\Lambda\to L}||e^{it(H_\Lambda+\sum_{a>0}B_{a\Lambda})}A
e^{-it(H_\Lambda+\sum_{a>0}B_{a\Lambda})}-\alpha^tA||=0\eqno{(4)}      $$
uniformly for $t$ in compact intervals of ${\bf R}$.}
\medskip
	We prove (4).  Write $\alpha_\Lambda^tA=e^{it(H_\Lambda+\sum_{a>0}B_{a\Lambda})}Ae^{-it(H_\Lambda+\sum_{a>0}B_{a\Lambda})}$ and $\delta_\Lambda A=i[H_\Lambda+\sum_{a>0}B_{a\Lambda},A]$.  If $A\in\cup_X{\cal A}_X$ we see using (1) that
$$   \alpha_\Lambda^tA=\sum_{m=0}^\infty{t^m\over m!}\delta_\Lambda^mA   $$
converges uniformly in $\Lambda$ for $|t|<\lambda/2(||\Phi||_\lambda+K)$.  We show in the lemma below that $\delta_\Lambda^mA\to\delta^mA$ in ${\cal A}$ when $\Lambda\to L$.  Therefore 
$$	\lim_{\Lambda\to L}||\alpha_\Lambda^tA-\alpha^tA||=0      $$ 
when $A\in\cup_X{\cal A}_X$, uniformly for $|t|\le T<\lambda/2(||\Phi||_\lambda+K)$.  But the condition $A\in\cup_X{\cal A}_X$ is removed by density, and the condition $|t|\le T<\lambda/2(||\Phi||_\lambda+K)$ by use of the group property.  The proof of (3) is similar.\qed
\medskip
	{\bf Lemma.}
\medskip
	{\sl If $A\in{\cal A}_\lambda$, then
$$	\lim_{\Lambda\to L}||\delta^mA-\delta_\Lambda^mA||=0      $$}
\indent
	From Chapter 2 Section 3 (last Proposition) we know that, if $\mu<\lambda$, $\delta^m$ maps ${\cal A}_\lambda$ into ${\cal A}_\mu$, and
$$	||\delta^m(A)||_\mu
\le||A||_\lambda m!(2(\lambda-\mu)^{-1}||\Phi||_\lambda)^m\eqno{(5)}      $$
\indent
	We write now  $\delta_\Lambda=\delta_\Lambda'+\delta_\Lambda''$, where
$$	\delta_\Lambda'A=i[H_\Lambda,A]\qquad,
	\qquad\delta_\Lambda''A=i[\sum_{a>0}B_{a\Lambda},A]      $$
Using (5) for $m=1$, and (1), we get
$$  ||\delta A||_\mu\le||A||_\lambda.2(\lambda-\mu)^{-1}||\Phi||_\lambda  $$
$$	||\delta_\Lambda'A||_\mu
	\le||A||_\lambda.2(\lambda-\mu)^{-1}||\Phi||_\lambda      $$
$$	||\delta_\Lambda''A||_\mu\le||A||_\lambda.2(\lambda-\mu)^{-1}K      $$
Given $\epsilon>0$ and $A\in{\cal A}_\lambda$ we can find $X$ such that $A=A_1+A_2$ with $A_1\in{\cal A}_X$ and $||A_2||_\lambda<\epsilon$.  Therefore 
$$	||(\delta-\delta_\Lambda)A||_\mu\le
	||(\delta-\delta_\Lambda)A_1||_\mu+||\delta A_2||_\mu
	+||\delta_\Lambda'A_2||_\mu+||\delta_\Lambda''A_2||_\mu      $$
$$	=||(\delta-\delta_\Lambda)A_1||_\mu
	+\epsilon.2(\lambda-\mu)^{-1}(2||\Phi||_\lambda+K)\eqno{(6)}      $$
Taking $\Lambda\supset\Lambda_X$ we also have  
$$	\delta_\Lambda''A_1=0      $$
by (2), and 
$$	(\delta-\delta'_\Lambda)A_1
	=i\sum_{Y:Y\not\subset\Lambda,Y\cap X\ne\emptyset}[\Phi(Y),A_1]      $$
so that
$$	||(\delta-\delta_\Lambda')A_1||_\mu
\le||A_1||_\lambda.2(\lambda-\mu)^{-1}||\Phi||'_{X\lambda}\eqno{(7)}      $$
where $||\Phi||'_{X\lambda}=\sup_{x\in X}\sum_{Y\ni x,Y\not\subset\Lambda}e^{({\rm card}Y-1)\lambda}||\Phi(Y)||$.  When $\Lambda\to L$ we have $||\Phi||'_{X\lambda}\to0$ and (6), (7) yield 
$$	\lim_{\Lambda\to L}||(\delta-\delta_\Lambda)A||_\mu=0\eqno{(8)}      $$
\indent
	We can now prove that, if $||\Phi||_\lambda<\infty$ and $A\in{\cal A}_\lambda$,
$$	\lim_{\Lambda\to L}||\delta^mA-\delta_\Lambda^mA||=0\eqno{(9)}      $$
We have indeed
$$	\delta^mA-\delta_\Lambda^mA
=\sum_{k=0}^{m-1}\delta_\Lambda^{m-k-1}(\delta-\delta_\Lambda)\delta^kA      $$
and, using (5), 
$$	||\delta^kA||_{2\lambda/3}
	\le||A||_\lambda.k!({6\over\lambda}||\Phi||_\lambda)^k      $$
hence, by (8),
$$  \lim_{\Lambda\to L}||(\delta-\delta_\Lambda)\delta^kA||_{\lambda/3}=0  $$
so that, using (5),
$$	||\delta_\Lambda^{m-k-1}(\delta-\delta_\Lambda)\delta^kA||
	=||\delta_\Lambda^{m-k-1}(\delta-\delta_\Lambda)\delta^kA||_0      $$
$$	\le||(\delta-\delta_\Lambda)\delta^kA||_{\lambda/3}(m-k-1)!
	({6\over\lambda}||\Phi||_\lambda)^{m-k-1}      $$
which tends to zero when $\Lambda\to L$.  This concludes the proof of (9).
\medskip
	{\bf 4. Nonequilibrium steady states.}
\medskip
	The interaction $\sum_{a>0}\beta_a\Phi_a$, evaluated at $X$ is $\beta_a\Phi_a(X)$ if $X\subset R_a$ and 0 if $X$ is not contained in one of the $R_a$.  The corresponding one-parameter group $(\beta^t)$ of automorphisms of ${\cal A}$ has, according to (A3), the KMS state\footnote{*}{The state $\sigma$ corres
ponds to the inverse temperature $+1$ rather than the inverse temperature $-1$ favored in the mathematical literature.} $\sigma=\otimes_{a\ge0}\sigma_a$ where $\sigma_0$ is the normalized trace on ${\cal A}_0={\cal A}_S$.  In fact 
$$	\sigma(A)=\lim_{\Lambda\to L}
{{\rm Tr}_{{\cal H}_\Lambda}(\exp(-\sum_a\beta_a(H_{a\Lambda}+B_{a\Lambda}))A)
	\over{{\rm Tr}_{{\cal H}_\Lambda}
	\exp(-\sum_a\beta_a(H_{a\Lambda}+B_{a\Lambda}))}}\eqno{(10)}      $$
\medskip
	{\bf Definition.}
\medskip
	We call {\it nonequilibrium steady states} (NESS) associated with $\sigma$ the limits when $T\to\infty$ of 
$$	{1\over T}\int_0^Tdt\,(\alpha^t)^*\sigma      $$
using the $w^*$-topology on the dual ${\cal A }^*$ of ${\cal A }$.  With respect to this topology, the set $\Sigma$ of NESS is compact, nonempty, and the elements of $\Sigma$ are $(\alpha^t)^*$-invariant states on ${\cal A }$.
\medskip
	{\bf Remark.} (Dependence on the decomposition $L=S+R_1+R_2+\ldots$)
\medskip
	Our definition of $\sigma$, and therefore of $\Sigma$ depends on the choice of a decomposition of $L$ into small system and reservoirs.  If $S$ is replaced by a finite set $S'\supset S$ and the $R_a$ by correspondingly smaller sets $R'_a\subset R_a$ one checks that (A1), (A2), (A3) remain valid.  If $\Phi'_a$ is the restriction of $\Phi$ to subsets of $R'_a$, the replacement of $\sum\beta_a\Phi_a$ by $\sum\beta_a\Phi'_a$ changes $(\beta^t)$ to a one-parameter group $(\beta'^t)$ and $\sigma$ to a state $\sigma'$.  The map $\sigma\to\sigma'$ (of KMS states for $(\beta^t)$ to KMS states for $(\beta'^t)$) is nonlinear (as can be guessed from (10)) and therefore we cannot expect that ${1\over T}\int_0^Tdt\,(\alpha^t)^*\sigma'$ has the same limit as ${1\over T}\int_0^Tdt\,(\alpha^t)^*\sigma$ in general, but the deviation is not really bad.  The (central) decomposition of KMS states into extremal KMS states gives factor states.  If $\sigma$ is assumed to be a factor state, and $(\alpha^t)$ is asymptotically abelian, one finds that $\lim{1\over T}\int_0^Tdt\,(\alpha^t)^*\sigma$ does not depend on the decomposition $L=S+R_1+R_2+\ldots$.  In other words:
\medskip
	{\sl Using the above notation, assume that $\sigma$ is a factor state, and that
$$	\lim_{t\to\infty}||[\alpha^tA,B]||=0      $$
when $A,B\in{\cal A}$.  Then, when $T\to\infty$, 
$$	\lim{1\over T}\int_0^Tdt\,(\alpha^t)^*\sigma'=
	\lim{1\over T}\int_0^Tdt\,(\alpha^t)^*\sigma      $$}
\medskip
	The proof is not hard.
\medskip
	{\bf 5. Entropy production.}
\medskip
	For finite $\Lambda\subset L$ we have defined 
$$	H_\Lambda=\sum_{X\subset\Lambda}\Phi(X)      $$
but $H_L$, $H_{R_a}$ do not make sense.  We can however define
$$	[H_L,H_{R_a}]=\lim_{\Lambda\to L}[H_\Lambda,H_{R_a\cap\Lambda}]
	=\lim_{\Lambda\to L}[H_\Lambda,H_{a\Lambda}]      $$
We have indeed
$$	[H_\Lambda,H_{a\Lambda}]
	=[H_\Lambda-H_{a\Lambda},H_{a\Lambda}]
	=[H_\Lambda-\sum_{b>0}H_{b\Lambda},H_{a\Lambda}]      $$
and (A2) gives
$$	H_\Lambda-\sum_{b>0}H_{b\Lambda}
	=\sum_{x\in S}\sum_{X:x\in X\subset\Lambda}
	{1\over{\rm card}(X\cap S)}\Phi(X)      $$
[implying the existence of the limit $\lim_{\Lambda\to L}(H_\Lambda-\sum_{b>0}H_{b\Lambda})=H_L-\sum_{b>0}H_{R_b}\in{\cal A}$].  Using (A1) we obtain
$$	||[\Phi(X),H_{a\Lambda}]||
\le2\lambda^{-1}||\Phi||_\lambda||\Phi(X)||e^{\lambda{\rm card}X}      $$
hence 
$$	\sum_{X\ni x}||[\Phi(X),H_{a\Lambda}]||
	\le2\lambda^{-1}||\Phi||_\lambda e^{\lambda}||\Phi||_\lambda      $$
and $[H_\Lambda,H_{a\Lambda}]$ has a limit $[H_L,H_{R_a}]\in{\cal A }$ when $\Lambda\to L$ with 
$$	||[H_L,H_{R_a}]||
	\le2{\rm card}S\lambda^{-1}e^\lambda||\Phi||_\lambda^2      $$
The operator 
$$	i[H_L,H_{R_a}]      $$
may be interpreted as the rate of increase of the energy of the reservoir $R_a$ or (since this energy is infinite) rather the rate of transfer of energy to $R_a$ from the rest of the system.  According to conventional wisdom we define the rate of entropy production in an $(\alpha^t)^*$-invariant state $\rho$ as 
$$	e_\rho=\sum_{a>0}\beta_a\rho(i[H_L,H_{R_a}])      $$
(this definition does not require that $\rho\in\Sigma$).
\medskip
	{\bf Remark.}
\medskip
	If we replace $S$ by a finite set $S'\supset S$ and the $R_a$ by the correspondingly smaller sets $R'_a\subset R_a$, we have noted earlier that (A1), (A2), (A3) remain satisfied.  As a consequence of (A1) we have 
$$	i[H_L,H_{R_a}-H_{R'_a}]
=\lim_{\Lambda\to L}i[H_\Lambda,H_{a\Lambda}-H'_{a\Lambda}]
=\lim_{\Lambda\to L}\delta(H_{a\Lambda}-H'_{a\Lambda})      $$
(where the operator $\delta$ has been defined just after (A3)), hence
$$	\rho(i[H_L,H_{R_a}-H_{R'_a}])=\lim_{\Lambda\to L}
	\rho(\delta(H_{a\Lambda}-H'_{a\Lambda}))=0      $$
{\it i.e.}, the rate of entropy production is unchanged when $S$ and the $R_a$ 
are replaced by $S'$ and the $R'_a$.  The reason why we do not have $\rho(i[H_L,H_{R_a}])=0$ is mathematically because $H_{R_a}$ is ``infinite'' ($H_{R_a}\notin{\cal A }$), and physically because our definition of $\rho(i[H_L,H_{R_a}])$ takes into account the flux of energy into $R_a$ from $S$, but not the flux at infinity.
\medskip
	{\bf Theorem} (see [4]).
\medskip
	{\sl The entropy production in a {\rm NESS} is nonnegative, {\it i.e.}, $e_\rho\ge0$ if $\rho\in\Sigma$.} 
\medskip
	We have seen that 
$$	[H_L,H_{R_a}]=\lim_{\Lambda\to L}[H_\Lambda,H_{a\Lambda}]      $$
$$	=\lim_{\Lambda\to L}
	[H_\Lambda-\sum_{b>0}H_{b\Lambda},H_{a\Lambda}]      $$
Therefore, using (A3) and $[H_{b\Lambda}+B_{b\Lambda},\sum_{a>0}\beta_a(H_{a\Lambda}+B_{a\Lambda})]=0$, we find
$$	\sum_{a>0}\beta_a[H_L,H_{R_a}]=\lim_{\Lambda\to L}
[H_\Lambda-\sum_{b>0}H_{b\Lambda},\sum_{a>0}\beta_aH_{a\Lambda}] $$
$$	=\lim_{\Lambda\to L}[H_\Lambda-\sum_{b>0}H_{b\Lambda},
	\sum_{a>0}\beta_a(H_{a\Lambda}+B_{a\Lambda})] $$
$$	=\lim_{\Lambda\to L}[H_\Lambda+\sum_{b>0}B_{b\Lambda},
	\sum_{a>0}\beta_a(H_{a\Lambda}+B_{a\Lambda})] $$
in the sense of norm convergence.
\medskip
	We also have, for some sequence of values of $T$ tending to infinity and all $A\in{\cal A}$,
$$	\rho(A)=\lim_{T\to\infty}{1\over T}\int_0^Tdt\,\sigma(\alpha^tA)
	=\lim_{T\to\infty}\lim_{\Lambda\to L}
	{1\over T}\int_0^Tdt\,\sigma(\alpha_\Lambda^tA)   $$
where, by (4),
$$	\alpha_\Lambda^tA=e^{it(H_\Lambda+\sum_{a>0}B_{a\Lambda})}
Ae^{-it(H_\Lambda+\sum_{a>0}B_{a\Lambda})}\to\alpha^tA\hbox{ in norm}      $$
when $\Lambda\to L$, uniformly for $t\in[0,T]$.
\medskip
	Write 
$$	H_{B\Lambda}=H_\Lambda+\sum_{a>0}B_{a\Lambda}      $$
$$	G_\Lambda=\sum_{a>0}\beta_a(H_{a\Lambda}+B_{a\Lambda})
	+\log{\rm Tr}_{{\cal H}_\Lambda}
	\exp(-\sum_{a>0}\beta_a(H_{a\Lambda}+B_{a\Lambda}))      $$
Then the entropy production is 
$$	e_\rho=\rho(i\sum_{a>0}\beta_a[H_L,H_{R_a}])
	=\lim_{T\to\infty}\lim_{\Lambda\to L}{i\over T}\int_0^Tdt\,
	\sigma(e^{itH_{B\Lambda}}
	[H_{B\Lambda},G_\Lambda]e^{-itH_{B\Lambda}})      $$
and the convergence when $\Lambda\to L$ of the operator $(e^{itH_{B\Lambda}}[H_{B\Lambda},G_\Lambda]e^{-itH_{B\Lambda}})$ is uniform for $t\in[0,T]$.  According to (A3) we may choose the $\Lambda$ tending to $L$ such that ${\rm Tr}_{{\cal H}_\Lambda}e^{-G_\Lambda}(\cdot)$ tends to $\sigma(\cdot)$ in the $w^*$-topology, hence
$$	e_\rho=\lim_{T\to\infty}\lim_{\Lambda\to L}{i\over T}\int_0^Tdt\,
	{\rm Tr}_{{\cal H}_\Lambda}(e^{-G_\Lambda}e^{itH_{B\Lambda}}
	[H_{B\Lambda},G_\Lambda]e^{-itH_{B\Lambda}})      $$
$$	=\lim_{T\to\infty}\lim_{\Lambda\to L}{1\over T}\int_0^Tdt\,
	{\rm Tr}_{{\cal H}_\Lambda}(e^{-G_\Lambda}{d\over dt}
	(e^{itH_{B\Lambda}}G_\Lambda e^{-itH_{B\Lambda}}))      $$
$$	=\lim_{T\to\infty}\lim_{\Lambda\to L}{1\over T}
	\big({\rm Tr}_{{\cal H}_\Lambda}(e^{-G_\Lambda}e^{iTH_{B\Lambda}}
	G_\Lambda e^{-iTH_{B\Lambda}})
	-{\rm Tr}_{{\cal H}_\Lambda}(e^{-G_\Lambda}G_\Lambda)\big)    $$
and the Theorem follows from the Lemma below, applied with $A=G_\Lambda$, $U=e^{iTH_{B\Lambda}}$ and $\phi(s)=-e^{-s}$.\qed
\medskip
	{\bf Lemma.}
\medskip
	{\sl Let $A$, $U$ be a hermitean and a unitary $n\times n$ matrix respectively, and $\phi:{\bf R}\to{\bf R}$ be an increasing function.  Then
$$	{\rm tr}(\phi(A)UAU^{-1})\le{\rm tr}(\phi(A)A)      $$}
\indent
	As R. Seiler kindly pointed out to me, this lemma can be obtained readily from O. Klein's inequality
$$	{\rm tr}(f(B)-f(A)-(B-A)f'(A))\ge0      $$
where $A$, $B$ are hermitean and $f$ convex: take $B=UAU^{-1}$ and $\phi=f'$.\qed
\medskip
	{\bf 6. The approach of Jak\v si\'c and Pillet.}
\medskip
	A more abstract approach to the positivity of entropy production is as follows [2]\footnote{*}{We have changed the notation of [2] to align it with the one used above.}.
\medskip
	We are given a C$^*$-algebra ${\cal A}$ with identity, an element $V=V^*\in{\cal A}$, time evolutions $(\breve\alpha^t)$, $(\alpha^t)$ ({\it i.e.}, strongly continuous one-parameter groups of $*$-automorphisms of ${\cal A}$) such that 
$$	\alpha^t(A)=\breve\alpha^t(A)+\sum_{n\ge1}i^n
	\int_0^tdt_1\int_0^{t_1}dt_2\ldots\int_0^{t_{n-1}}dt_n
	[\breve\alpha^{t_n}(V),[\ldots[\breve\alpha^{t_1}(V),A]]]      $$
and an $(\breve\alpha^t)$-invariant state $\sigma$ on ${\cal A}$.  Therefore $(\alpha^t)$ is a local perturbation by $V$ of the ``free'' evolution given by $(\breve\alpha^t)$ -- see Section 4 of Chapter 2 -- and $\sigma$ is an invariant state for the ``free'' evolution.  We furthermore assume that
\medskip
	(C1) There exists a time evolution $(\beta^t)$ for which $\sigma$ is a KMS state at inverse temperature $+1$
\medskip
	(C2) $V$ is in the domain of the infinitesimal generator of $(\beta^t)$.
\medskip\noindent
[In fact Jak\v si\'c and Pillet assume a temperature $-1$ in (C1); our choice of temperature $+1$ will bring a change of sign below in the definition of the entropy production.  In the situation discussed in Sections 2-5 we have 
$$	V=\sum_{X\cap S\ne\emptyset}\Phi(X)      $$
hence $||V||_\lambda\le||\Phi||_\lambda{\rm card}S$, and $V\in{\cal A}_\lambda$.  Note that ${\cal A}_\lambda$ is in the domain of the infinitesimal generator $\delta_\beta$ of $(\beta^t)$ by Section 3 of Chapter 2, hence (C2) holds.  The advantage of the approach of Jak\v si\'c and Pillet is that $\sigma$ can be an arbitrary KMS state: the existence of ``boundary terms'' $B_{a\Lambda}$ such that (10) holds is not required].
\medskip
	In this setup one introduces the observable
$$	-\delta_\beta(V)      $$
and the {\it entropy production} in the state $\rho$ is defined as 
$$	\rho(-\delta_\beta(V))      $$
[In the situation discussed in Section 5 we have 
$$	-\delta_\beta(V)=-\sum_{a>0}\beta_a\sum_{X\subset R_a}
	\sum_{Y:Y\cap S\ne\emptyset}i[\Phi(X),\Phi(Y)]      $$
$$	=\sum_{a>0}\beta_ai[H_L,H_{R_a}]      $$
so that $\rho(-\delta_\beta(V))=e_\rho$ is indeed the rate of entropy production in the state $\rho$].
\medskip
	{\bf Finite dimensional digression.}
\medskip
	For the purpose of motivation we discuss now the case where ${\cal A}$ would be the algebra of $n\times n$ matrices, and consider two states on ${\cal A}$ given by density matrices $\mu$, $\nu$.  A relative entropy is then defined by 
$$	{\rm Ent}(\mu|\nu)=-{\rm tr}(\mu\log\mu-\mu\log\nu)\le0      $$
If $(\alpha^t)$ is a one parameter group of $*$-automorphisms of ${\cal A}$ we have thus 
$$	{d\over dt}{\rm Ent}(\mu\circ\alpha^t|\nu)
	={\rm tr}(\mu{d\over dt}\alpha^t(\log\nu))      $$
Suppose now that $\nu$ is preserved by the ``free'' evolution $(\breve\alpha^t)$, and that $(\alpha^t)$ is a perturbation of $(\breve\alpha^t)$, so that 
$$	\alpha^t(A)=e^{i(H+V)t}Ae^{-i(H+V)t}\qquad,\qquad
	\breve\alpha^t(A)=e^{iHt}Ae^{-iHt}      $$
then
$$	{d\over dt}\alpha^t(\log\nu)=\alpha^t(i[V,\log\nu])      $$
Define now $(\beta^t)$ by 
$$	\beta^t(A)=e^{-it\log\nu}Ae^{it\log\nu}      $$
so that $\nu$ is the corresponding KMS state (at inverse temperature $+1$).  Then if $\delta_\beta$ is the infinitesimal generator of $(\beta^t)$ we have 
$$	i[V,\log\nu]=\delta_\beta(V)      $$
hence 
$$	{d\over dt}\alpha^t(\log\nu)=\alpha^t(\delta_\beta(V))      $$
$$	{d\over dt}{\rm Ent}(\mu\circ\alpha^t|\nu)
	=\mu(\alpha^t(\delta_\beta(V)))      $$
We obtain thus
$$	{\rm Ent}(\mu\circ\alpha^T|\nu)-{\rm Ent}(\mu|\nu)
	=\int_0^T(\mu\circ\alpha^t)(\delta_\beta(V))\,dt      $$
or, taking $\mu=\nu=\sigma$,
$$	0\le-{\rm Ent}(\sigma\circ\alpha^T|\sigma)
	=\int_0^T(\sigma\circ\alpha^t)(-\delta_\beta(V))\,dt      $$
\indent
	{\bf The infinite dimensional situation.}
\medskip
	If $\mu$, $\nu$ are two faithful normal states on a von Neumann algebra ${\cal M}$ [in our case $\pi_\sigma({\cal A})''$], Araki has introduced a relative entropy ${\rm Ent}(\mu|\nu)$ in terms of a relative modular operator associated with $\mu$, $\nu$.  We must refer the reader to [1] Definition 6.2.29 for details.  Using this definition, Jak\v si\'c and Pillet have worked out an infinite dimensional version of the finite dimensional calculation given above.  They are able to prove the formula
$$	\int_0^T(\sigma\circ\alpha^t)(-\delta_\beta(V))\,dt
	=-{\rm Ent}(\sigma\circ\alpha^T|\sigma)\ge0      $$
which can be interpreted as an entropy balance, and gives in the limit
$$	\rho(-\delta_\beta(V))\ge0      $$
if $\rho$ is a NESS.  We shall not go into the details of the proof, which is relatively technical.
\medskip
	The approach of Jak\v si\'c and Pillet has the interest of great generality.  In particular $\sigma$ can be an arbitrary KMS state.  Also, instead of a spin lattice system one can consider fermions on a lattice.  For a noninetacting fermion model, Jak\v si\'c and Pillet have announced a proof of strict positivity of the entropy production, as had been suggested in [4].
\medskip
	{\bf 7. Further work.}
\medskip
	The assumptions (A1), (A2), (A3) or (C1), (C2) are quite weak (and thus relatively easy to check).  They allow the definition of nonequilibrium steady states (NESS) and a proof that the entropy production is $\ge0$.  Other questions arise naturally: is there a unique NESS $\rho$ associated with $\sigma$?  How does $\rho$ vary with the interaction $V$ (linear response)?  These questions are tackled in [3] under strong conditions of asymptotic abelianness in time.  Unfortunately, these conditions appear very difficult to verify, and that seems to be a major obstacle on our way to understanding nonequilibrium for quantum systems.
\medskip
	{\bf References.}

[1] O. Bratteli and D.W. Robinson.  {\it Operator algebras and quantum statistical mechanics I, II.}  Springer, New York, 1979-1981.  [There is a 2-nd ed. (1997) of vol. II].

[2] V. Jak\v si\'c and C.-A. Pillet.  ``On entropy production in quantum statistical mechanics.''  Preprint.

[3] D.Ruelle.  ``Natural nonequilibrium states in quantum statistical mechanics.''  J. Statist. Phys. {\bf 98},57-75(2000).

[4] D.Ruelle.  ``Entropy production in quantum spin systems.''  Preprint.
\end